\documentclass[preprint,showpacs,preprintnumbers,amsmath,amssymb,superscriptaddress]{revtex4}

\usepackage{graphicx}
\usepackage{dcolumn}
\usepackage{bm}

\begin{document}

\title{Some remarks about the $\beta$-delayed $\alpha$-decay of $^{16}$N}

\author{L. Buchmann}
\email {lothar@triumf.ca}
\affiliation {TRIUMF, 4004 Wesbrook Mall, Vancouver, British Columbia, Canada, V6T 2A3}
\author{G. Ruprecht}
\affiliation {TRIUMF, 4004 Wesbrook Mall, Vancouver, British Columbia, Canada, V6T 2A3}
\author{C. Ruiz}
\affiliation {TRIUMF, 4004 Wesbrook Mall, Vancouver, British Columbia, Canada, V6T 2A3}

\begin{abstract}
The $\beta$-delayed $\alpha$-decay of $^{16}$N has been used to restrict the 
E1 fraction of the ground state $\gamma$-transition 
in the astrophysically important 
$^{12}$C($\alpha$,$\gamma$)$^{16}$O reaction in several experiments including
those performed at TRIUMF and several other laboratories. 
A review of the published measurements is given,
and GEANT4 simulations and  R-Matrix calculations are 
presented to further clarify the observed $\alpha$ spectra. 
A clear response-function, in the form of a low-energy tail from the scattering 
of $\alpha$-particles in the catcher foil is
observed in these simulations for any foil thickness.
Contrary to claims in the literature,
the simulations show that the TRIUMF measurement and those performed at Yale
and Mainz originate from the same underlying spectrum.
The simulations suggests that the discrepancies between 
the Yale and TRIUMF final results can be
attributed to incorrect deconvolution methods applied in the former
case. 
The simulations show in general that the form (width) of the spectrum is very
sensitive to the catcher foil thickness.
It is concluded that the TRIUMF measurement
most likely represents the currently closest approximation to the true 
$\beta$-delayed $\alpha$-decay spectrum of $^{16}$N.
\end{abstract}


\maketitle

\section {Introduction}

The $\beta$-delayed $\alpha$-decay spectrum of $^{16}$N has been used to restrict the
$E1$ fraction of the ground state transition in 
$^{12}$C($\alpha$,$\gamma$)$^{16}$O. This reaction
has often been referred to as the ``holy grail" of experimental 
Nuclear Astrophysics due both to the importance of the
$^{12}$C($\alpha$,$\gamma$)$^{16}$O reaction in stellar evolution,
and the experimental difficulty in measuring the cross section at low
enough energies 
to be relevant for quasi static helium burning. Some possible 
assistance in determining the 
$E1$ ground state transition rate arises
from the fact that, in the $\beta$-delayed $\alpha$-decay of $^{16}$N, the subthreshold 
E$_x$=7.117, J$^{\pi}$=1$^-$ state of $^{16}$O is populated with a yield of 
about 0.05 in the $\beta$-decay
of $^{16}$N, while the unbound E$_x$=9.60 MeV, J$^{\pi}$=1$^-$ state is 
populated with a yield of about
10$^{-5}$. This leads to a relatively stronger influence of the subthreshold
state on the $\beta$-delayed $\alpha$-decay spectrum of $^{16}$N, particularly
on its low energy part. 
As the ground state of $^{16}$N has a spin and
parity of J$^{\pi}$=2$^-$, only $^{16}$O states with 
J$^{\pi}$=1$^-$ and 3$^{-}$ will contribute to allowed $\beta$ $\alpha$ decay of $^{16}$N, with
no J$^{\pi}$=3$^-$ state being nearby the given energy range of the aforementioned
$\alpha$ spectrum. Therefore the J$^{\pi}$=3$^-$ contribution to the spectrum is 
expected to be small,
but not necessarily entirely negligible, while the relevant J$^{\pi}$=1$^-$ part is dominant.

While there is, of course, a true, underlying $\beta$-delayed $\alpha$-decay 
spectrum of $^{16}$N, any measurement of the spectrum
will lead to experiment-specific distortions on the spectrum, 
like those from detector resolution, collector foil effects
and possible backgrounds. While the last of these
can be removed by a correct subtraction, it is 
quite difficult to remove the former effects as they convolute the ideal spectrum with 
a usually energy-dependent response function. 
In this case each bin point of the measured spectrum will be mixed 
more or less with contributions from elsewhere in the spectrum. In general, 
deconvolutions\footnote {We use the expressions `convolution' and
`folding' (and their opposites) with the same meaning. However, in literature
often a distinction between `folding' with a response function $G(E-E')$
and `convolution' with a response function $G(E,E-E')$ is made.} 
are considered a difficult mathematical problem with uncertainties
growing rapidly the more iteration steps are applied and typically 
they involve some judgment calls. We will show that minimizing the
system response is the only practical solution and that all
attempts of a deconvolution lead to failure.

As response functions (likely) differ in technically different experiments,
experimental spectra of the $\beta$-delayed $\alpha$-decay 
of $^{16}$N may in principle
look rather different in comparison without implying 
a major difference in the underlying spectrum.
In a correct fitting procedure,
a theoretical curve should be convoluted with 
an experimentally well known response function,
and then be compared with the data; the theoretical curve can then be 
varied to fit the data after  
convolution. The theoretical fits may then be compared among different
experiments. Of course, the sensitivity and the response of
an experiment should be estimated beforehand independent of the
actual $^{16}$N spectrum.

There have been several measurements of the 
$\beta$-delayed $\alpha$-decay spectrum of $^{16}$N
\cite {Waf71,Azu94,Zha93,Fra97,Fra96,Fra07,Ade94,Tan07}. These measurements will be 
discussed below and compared, with the measurement 
described in Ref. \cite {Azu94} serving as a benchmark. In Sec. 
\ref {sec:discmeas} we will describe individual measurements and their
respective methods to obtain their final results, in Sec. \ref {sec:geant}
we will present GEANT IV simulations of the experiments, in Sec. \ref {sec:compmeas}
different measurements are compared to each other, in Sec. \ref {sec:conclusion}
the final conclusions will be drawn and in Sec. \ref {sec:outlook}
some future possibilities regarding the 
$\beta$-delayed $\alpha$-decay spectrum of $^{16}$N will be discussed. 
In particular, besides many
other issues of more minor concern, it will be shown that
\begin {itemize}
\item the spectrum obtained at Mainz University 
                   suffers from considerable distortion effects;
\item the method applied to deconvolute the spectra obtained at Yale University 
             appears to be mathematically erroneous;
\item the measurement done in Argonne National Laboratory 
                       likely suffers from an unreported background.
\end {itemize}

\section {Discussion of measurements}
\label {sec:discmeas}

\subsection {The Mainz spectrum}

In the late 1960's to early 1970's a group at Mainz University\footnote {We label
the experiments subsequently by the laboratory where they took place.}
led by H. W\"{a}ffler
made a series of very high statistics
measurements of the $\beta$-delayed $\alpha$ spectrum of $^{16}$N 
\cite {Hat69,Hat70,Neu74} 
 with the ultimate goal of detecting the parity-forbidden
$\alpha$ decay of the E$_x$=8.872 MeV J$^{\pi}$=2$^-$ state in $^{16}$O.
A gas cell containing $^{15}$N gas was bombarded with a deuteron beam, producing
$^{16}$N via the (d,p) reaction. 
The activated gas was pumped into (one or) two detection cells, where the 
$\alpha$ decay was registered by (four or) 
eight silicon detectors mounted on the walls of the (one or) two counting cells. 
The silicon detectors were separated from the 6-8 Torr gas volume 
by 30 $\mu$g/cm$^2$ collodion (C$_6$H$_7$N$_{2.5}$O$_{10}$)
foils. No coincidence requirements were placed on the events to prevent
detection of $\alpha$-particles possibly degraded by system response.

\begin{figure}
\center
\hspace*{-0.5cm}
\includegraphics[width=8.5cm]{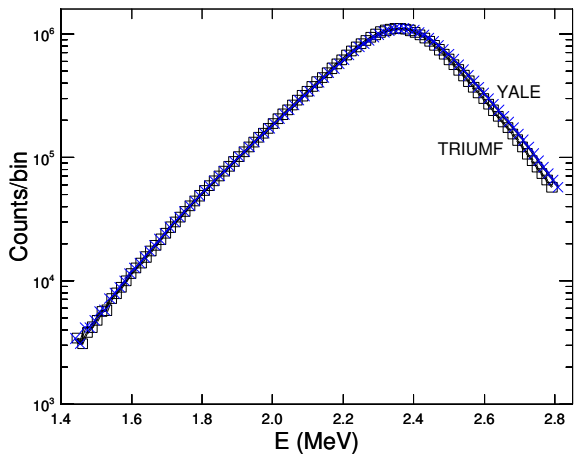}
\caption{(color online)
Comparison of the Mainz spectrum as calibrated in Ref. \cite {Fra07} (cross)
and Ref. \cite {Azu94} (box); the energy is displayed in the CM system.}
\label {fig:waeffler}
\end {figure}

In 1971 H. W\"{a}ffler communicated about a quarter of the events from the 
then available spectrum \cite {Hat70} to both C.A Barnes (Caltech) and  F. Barker 
(Australian National University) (about 3.2$\times$10$^7$ events). 
These letters contained a spectrum as 
counts/bin sorted by channel, and a very precise energy calibration 
of incompletely explained origin. The data sent were selected
for the smallest $\beta$-induced tail of low energy pulses, for which
the silicon detector biases were lowered to 8 V to reduce
the depletion depth in the silicon. It was confirmed, however, by H. W\"{a}ffler and
by the position of the $\alpha$ peak from 
the singly forbidden $\beta$-decay to the E$_x$=9.85 
J$^{\pi}$=2$^+$ $^{16}$O state in this spectrum that no correction for the 30 $\mu$g/cm$^2$
collodion foils has been included in the calibration \cite {Bar99,Waf99}. 
The matter was
also discussed in Ref. \cite {Buc06}. This spectrum, though unpublished,
has been widely used in comparison to later measurements of 
the $\beta$-delayed $\alpha$ spectrum of $^{16}$N (see, e.g. Ref. \cite {Tan07}) .
In particular, in Ref. \cite {Fra07} a table of an 
energy calibrated spectrum of the Mainz data
is published (Tab. A.1) without an adjustment for the foil thickness\footnote
{There is an additional error in Table A.1\cite {Fra07} as the points at 1468 and 1482 keV
have the same number of counts. The correct number of counts of the point at 1468 keV
should be 3785.}. Fig. \ref {fig:waeffler} shows
a comparison between the two calibrations.

\subsection {Spectra from the Yale group}

\subsubsection {France III et al. publications}
\label {sec:france}

As there is a recent publication by the Yale/U. of Conn. group \cite {Fra07} about their
second measurement of the $\beta$-delayed $\alpha$ spectrum of $^{16}$N,  this
work will be discussed here first among the two Yale measurements. 
However, the first measurement of this group
\cite {Zha93}, published in 1993, 
had basically the same features as the second experiment
with the exception of a different
data reduction from the raw ($\beta$-efficiency corrected) spectrum
to the final spectrum, and, as a result, a vastly different 
final spectrum (Sec. \ref {sec:compyale}).

\begin{figure}
\center
\includegraphics[width=10cm]{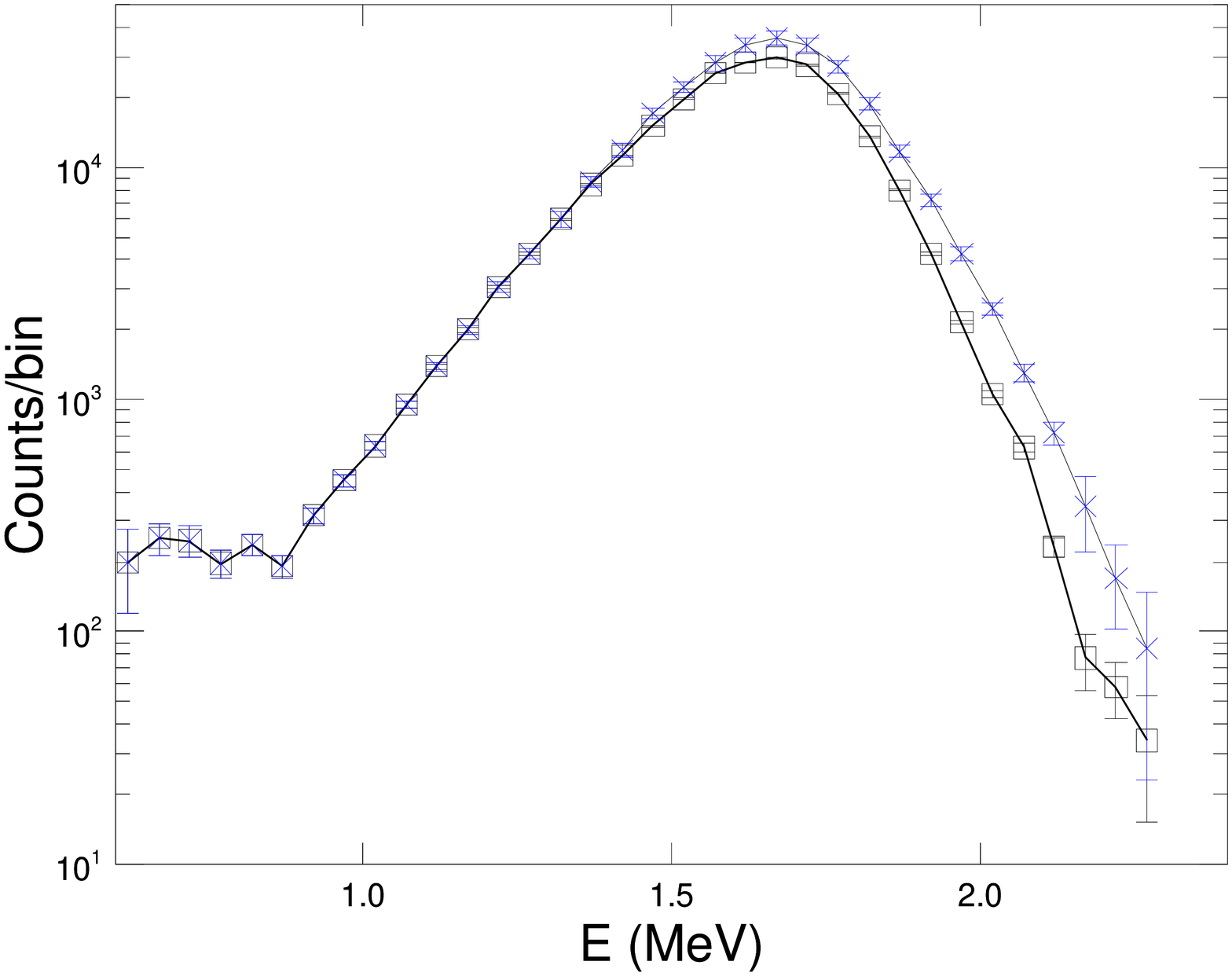}
\includegraphics[width=10cm,]{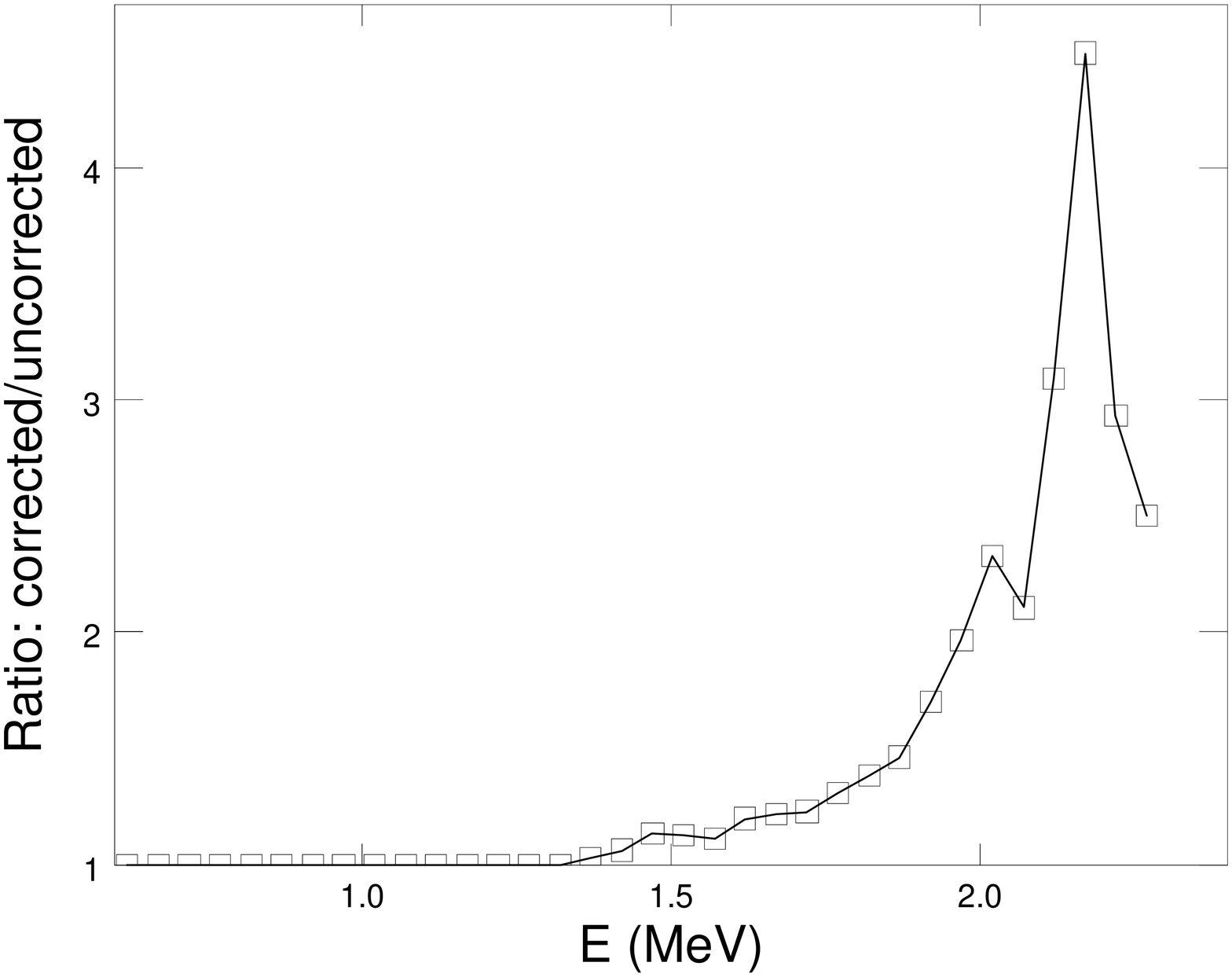}
\caption{(color online) Upper Panel: Comparison between the measured 
spectrum in Refs. \cite {Fra96,Fra07}(box) 
         and the $\beta$-corrected spectrum (cross). 
         Lower Panel: Ratio of the two spectra. The energy is in the laboratory system.}
\label {fig:france-nobeta}
\end {figure}

In both measurements the d($^{15}$N,p)$^{16}$N reaction was 
used for production with the $^{16}$N escaping the deuterium 
target at high velocity. To catch 
the $^{16}$N  nuclei from the low energy part of the recoiling $^{16}$N velocity 
distribution a tilted aluminum foil of 180 $\mu$g/cm$^2$
thickness was used. After the collection period, the catcher foil was moved in
front of a silicon counter array with $\beta$-counters positioned behind the foil. 
A $\beta$-$\alpha$ coincidence was used to discriminate against some kinds
of background events. To obtain the spectrum that was deconvoluted later,
a $\beta$-efficiency correction was necessary, particularly for the higher
energy $\alpha$ events where the coincident $\beta$ energies are very low. 
Fig. \ref {fig:france-nobeta}
shows a comparison between the measured and the $\beta$-efficiency-corrected
$\alpha$ spectrum and the ratio of the two spectra.
The correction rises with increasing energy 
to a factor of 4.5 and then falls for the highest energy point 
to about 2.5. Little is said in Refs.
\cite {Fra96,Fra07} how this correction for the extremely low energy $\beta$ particles 
at the high $\alpha$-energy side was actually derived. With decreasing $\beta$ energy 
one would expect, in general, a smoothly rising ratio, i.e. the inverted efficiency curve,
and a possible detection threshold for the $\beta$-rays.

\begin{figure}
\center
\includegraphics[width=6.5cm,angle=-90]{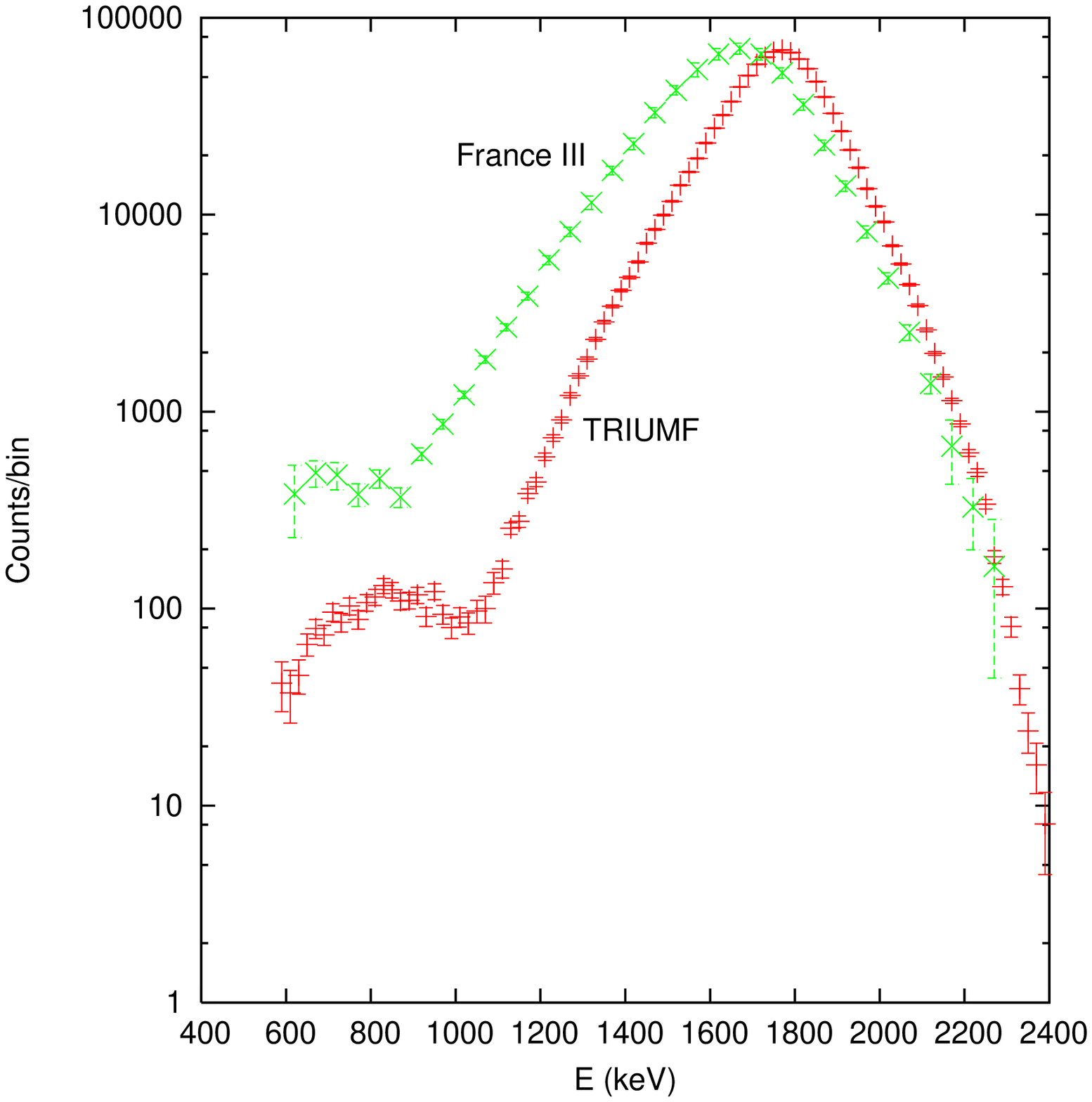}
\caption{(color online) Comparison of the final TRIUMF \cite {Azu94} (bars) 
and the $\beta$-efficiency corrected
$\alpha$-spectrum of Ref. \cite {Fra96} (crosses). 
The spectrum of Ref. \cite {Fra96} has been normalized
to the height of the TRIUMF spectrum. The abscissa represents the
$\alpha$-energy in the laboratory.}
\label {fig:trifra}
\end {figure}

About 235,000 events were produced \cite {Fra96,Fra07} from the analysis of the
time versus energy spectrum. The raw $\alpha$ spectrum ($\beta$-efficiency corrected,
285,000 events) from this analysis 
(displayed in Ref. \cite {Fra96}, Fig. 5.5) is shown in Fig.\ref {fig:trifra} in comparison
with the published TRIUMF spectrum \cite {Azu94}. 
It is quite obvious from Fig. \ref {fig:trifra} 
that the spectrum of Ref. \cite {Fra96,Fra07} is both shifted and broadened 
compared to the TRIUMF spectrum. Also the low energy plateau
is about an order of magnitude higher than in the TRIUMF measurement.
This is clearly an effect of the very thick collector
foil compared with the one used at TRIUMF (10 $\mu$g/cm$^2$ of carbon). In principle, a
careful simulation of the spectrum in Ref. \cite {Fra96,Fra07} 
might have produced information about the underlying $\beta$-delayed $\alpha$ spectrum,
though likely with a large uncertainty and considerable dependence
on the model used, see Sec. \ref{sec:yaledecon}. 
However, in Refs. \cite {Fra96,Fra07} a different approach to 
the data reduction was chosen, described  as follows:
\begin {quotation}
`The measured spectral line shape was corrected for distortions caused by the variability 
of our time and energy resolution, see Fig. 5. For a spectrum constant in energy the
yield measured at each point in that spectrum is directly proportional to the energy 
integration interval. This conclusion holds for a spectrum in any physical
variable. In the case of this experiment, the data are integrated over time 
and energy with the integration intervals being the time and energy resolutions. 
[...]\footnote {[...] notes here and in other quotations an omission of text
for clarification.}
The final spectrum, with these resolutions divided out, is shown in Fig. 6 [....].'
\end {quotation}
Note that no reference to an experiment is given, in which a similar deconvolution method
has been used.

The word `resolution' in regard to the energy is used here in a somewhat unusual way,
as it clearly labels the energy-dependent energy loss through the aluminum foil. 
The same is true for the `time resolution' which is just the time
of flight between the collector foil and the $\alpha$-particle detector, and not
the experimental spread of this time-of-flight signal.
The final spectrum shown in 
Refs. \cite {Fra97,Fra96,Fra07} is indeed derived by division
by this `energy resolution' and this `time resolution'.
Questions of dimensionality, i.e. divisions by energy and time,
and the obvious renormalization that must have been performed
to conserve the total number of counts are not discussed in these references.

The claim that an experimental
convolution, i.e. the result of a bad resolution, 
can be undone by dividing by an arithmetic function\footnote{Of course,
if the final result is known, it is possible to construct a
response function $G(E,E'-E)=\frac{f_f(E)}{f_i(E)}\delta(E'-E)$ with
$f_i(E)$ the initially measured spectrum and $f_f(E)$ the `unfolded' spectrum,
see Sec. \ref {sec:diszhao}. Such a function, however, is not independent
of the initial ($\alpha$) spectrum.} that is independent of the final spectrum
is not correct.
Such a general claim can be easily refuted by a counter-example.
%
In many cases in experimental
physics a $\delta$-like function (e.g. a narrow line) is folded 
in first order due to statistical processes in the event collection
by a Gaussian convolution.
No division by a finite number will recover the initial $\delta$-like
function from the Gaussian measured. 
The number of events will stay conserved by the convolution. 
If only a fraction of the Gaussian function is integrated,
the number of counts will, of course, change depending on the integration interval.
Such a change is neither linear, nor will the total number of counts ever be exceeded,
however far one may choose the integration interval. A reasonable integration interval
for such a peak will, of course, depend on the resolution, but is not identical to it.
In fact, by this method of division, 
for a $\delta$-shaped input function, a $\delta$-shaped output
function results as the input only needs to be multiplied
by the constant resolution.
I.e. no degradation or convolution of such a line will result.
It may be argued that the situation is different for a continuous
spectrum, where, indeed, the resolution may be energy dependent. However,
that would make the response function dependent on the spectral form, which
is not expected from a reasonable experiment.

\begin{figure}
\center
\includegraphics[width=6.5cm,angle=-90]{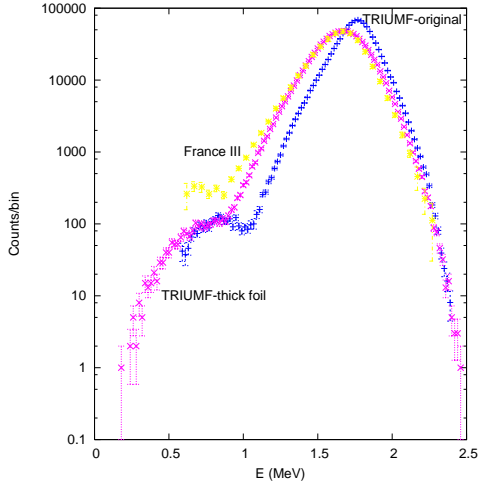}
\caption{(color online) A Monte Carlo simulation (1000000 events) 
of the $\beta$-delayed $\alpha$-decay spectrum of $^{16}$N
as found in Ref. \cite {Azu94} (bar points) exposed to the
energy loss as described in Refs. \cite {Fra96,Fra07} 
(cross points), and a comparison to the
France III raw spectrum (asterisk points), appropriately matched in height.}
\label {fig:n16stopfrance}
\end {figure}

It is straightforward to simulate, 
how the bare energy loss from the Yale catcher foil
will influence the $\beta$-delayed $\alpha$-decay spectrum of $^{16}$N. 
The results of the
Monte Carlo simulations are shown in Fig. \ref {fig:n16stopfrance}.
It is obvious that the energy loss effect gives a good description of the high
energy side of the France III spectrum but misses further response effects
on the low energy side (see Sec. \ref {sec:geant}). Therefore there 
is more to the response function than the simple stopping
power effect (see Sec \ref {sec:simyale}). 

The main contribution to the deconvolution by division in Ref. \cite {Fra07},
already shown to be incorrect, comes from the
time of flight resolution as displayed in their Fig. 5. as it shows a
steep energy dependence. This use of the time-of-flight spectrum
has no physical justification, as  (i) the time-of-flight measurement does not 
influence the pulse height in the energy spectrum;
(ii) the time-of-flight through the foil is about 50 fs for a 1 MeV $\alpha$-particle, 
therefore the implantation (start)
position in the catcher foil has nothing to do with the observed time-of-flight.
The increase in the time of flight for lower energy $\alpha$-particles 
should follow $1/\sqrt{E}$ which is clearly exceeded in Fig. 5 of Ref. \cite {Fra07}.
However, as the spectrum discussed in Refs. 
\cite {Fra96,Fra97,Fra07} is not derived from the time-of-flight information, but from the
pulse height in the silicon detectors, the time-of-flight, not influenced 
by the catcher foil thickness, has nothing to do with the response to the foil,
but only with some kinds of background discrimination applied. 
For example, events degraded in the silicon detectors will be removed by the time-of-flight 
information.

\begin{figure}
\center
\includegraphics[width=10cm]{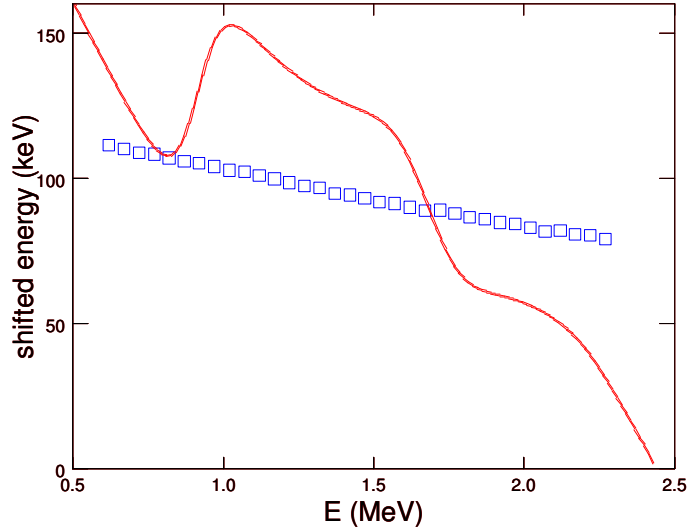}
\caption{(color online) Expected energy shift for the foil used in
Refs. \cite {Fra96,Fra07} (line) using Eq. \ref {eq:eeff}. The
original and the shifted spectrum are
shown in Fig. \ref {fig:n16stopfrance.}. 
In comparison, the energy shift used in Refs. \cite {Fra96,Fra07}
is shown here (bars). The energy is in the laboratory system.}
\label {fig:eshift}
\end {figure}

Besides a correction by division in the yield, the position of the energy points
(initial bins) has also been changed, 
i.e. shifted to higher energies. Of course, a correct deconvolution
does cause a shift and yield a change of channels simultaneously, see Sec. \ref {sec:yaledecon}. 
This is claimed to be done in the following way
\cite {Fra07}:
\begin {quotation}
`The effective energy of the emerging $\alpha$-particles for each data point was calculated 
using the {\it expected variation}\footnote {Highlighted here.} of the yield over the
energy width of the catcher foil for each slice. [...] Note that due to fast variation 
in the yield, the effective $\alpha$-particle energy is not the 
one due to $\alpha$-particles 
emitted from the center of the catcher foil.'
\end {quotation}
This remark appears to suggest that a preconceived knowledge of the final spectrum  
has been used to derive these corrections. 
It can now be checked, how far the energy shifts used in Refs. 
\cite {Fra97,Fra07} are consistent with the spectra shown in 
Fig. \ref {fig:n16stopfrance}. To calculate the effective energy 
$E_{eff}$ we solve numerically the integral
\begin {equation}
E_{eff}=\frac{\int_{E-\Delta(E)}^{E} E' w(E') dE'}{\int_{E-\Delta(E)}^{E} w(E') dE'}
\label {eq:eeff}
\end {equation}
for the effective energy. Here $E$ is the initial energy, $\Delta$
the foil thickness, and $w(E)$ the $\beta$-delayed $\alpha$-distribution of $^{16}$N
using $R$-matrix theory applied to the TRIUMF spectrum. For
simplification, the minor change in stopping power 
over the integration integral is being ignored. To calculate the energy shift,
we subtract the effective energy from the initial energy. 
Fig. \ref {fig:eshift} displays the
energy shift found here and the one used in Refs. \cite {Fra96,Fra07}.
Compared to the Monte Carlo calculations, taking also the energy dependence
of the stopping power into account, the curve shown in Fig. \ref {fig:eshift}
slightly underestimates the energy shift encountered.
While the absolute position of these curves on the abscissa may be disputed, 
it is clear that the energy dependence given by the calculation shown 
differs greatly from the
approximately linear relation used in Refs. \cite {Fra96,Fra07}. 
The use of the France III final spectrum as input would change little, if employed
in the simulations. In fact, it is the quickly changing yield over the energy 
loss range in the foil of
the $\beta$-delayed $\alpha$ spectrum that causes these quite distinct changes in the energy
correction. It is apparent that the energy correction used in Refs. \cite {Fra96,Fra07}
is equivalent to taking the
center of the catcher foil employing the nearly linear energy dependence
of the stopping power despite the opposite claim in the above quotation. 
This leads on the low energy side of the spectrum to
an underestimate of the necessary energy correction in Refs. \cite {Fra96,Fra07}, 
while the correction is considerably overdone on the high energy side.
Most importantly, calculating `the effective $\alpha$-particle energy' requires
a knowledge of the true distribution beforehand.

\subsubsection {Discussion of Z.Zhao et al.}

\label {sec:diszhao}

As mentioned above, in a measurement previous to Refs. \cite {Fra96,Fra07}
a $\beta$-delayed $\alpha$-spectrum was obtained at Yale University by largely the
same group \cite {Zha93}. The experimental method was
approximately identical to the one described for France III, except that the 
$\alpha$ collection system was changed from upstream
to downstream of the target. Therefore, a spectrum very similar to the
one shown for France III was collected (See Fig. 1a of Ref. \cite {Zha93}). 
However, the deconvolution method
was different. As this measurement has not been indicated as incorrect or superseded
in the later publications by this group and has been used for
comparison in a recent publication \cite {Tan07}, it will be discussed in the
subsequent section.

In Ref. \cite {Zha93} an $R$-matrix description of the spectrum measured is 
given, following the parametrization of Ref. \cite {Ji90}. Beyond this, 
the authors present in their Fig. 1b the ``$\ell$=3 component of our fit"
and conclude that ``only with the introduction of the small $\ell$=3 component
was it possible to reproduce the line shape of the unfolded spectrum".
It has since been stated \cite {Zha94} that the $\ell$=3 component of 
Ref. \cite {Zha93} is not the result of a fit, but simply ``ad-hoc'', i.e.
that it is an invented assumption by the authors without any basis
in theory\footnote{It may be noted that any attempt to find an R-matrix solution
based on three states for the $f$-wave presented in Ref. \cite {Zha93} leads to
no acceptable result.}. 

To see, however, the results of a consistent analysis of
the data of  Ref. \cite {Zha93} we have
analyzed those data using a full $R$-matrix fit, following the description of 
Ref. \cite {Azu94}. For the best fit a value of $\chi^2$/point=0.17 was found for the
data of Ref. \cite {Zha93}. This unexpectedly low value is attributed mainly to the
data points of the main peak of Fig. 1(b) of \cite {Zha93}
which show smaller fluctuations than those of Fig. 1(a), a result not normally
expected from the deconvolution of an experimental spectrum (see Sec. \ref {sec:yaledecon}). 
To further understand these findings, the deconvolution process
based on the thesis of Z. Zhao \cite {Zha93b} will be subsequently discussed and the likely
reason for the low $\chi^2_{\nu}$ presented.

The first step of the data reduction, 
after a $\beta$-efficiency correction (Sec. \ref {sec:france}), was to correct for the
energy loss in the target thickness. For this, it appears that the maximum of the experimental 
spectrum was
shifted to match the maximum of the Mainz spectrum without further corrections.

Independent of the energy shift, the number of counts in each bin has also undergone 
a data reduction procedure. As stated in Ref. \cite {Zha93}, 
the final result is {\it expected
to be similar} to that of Ref. \cite {Waf71}, which is referred to as the ``zero target
thickness" spectrum\footnote {The Mainz measurement \cite {Waf71} certainly
employed a larger catcher thickness than the TRIUMF\cite {Azu94} one, but none
of the experiment had zero target thickness, see Sec. \ref {sec:mainzsim}.}. 
Since the spectrum measured was far from agreeing with the one
of Ref. \cite {Waf71}, the authors chose to complete the unfolding of the data
by finding a response function which is to be 
determined ``semi-empirically'' \cite {Zha93}.

A response function, $[G(E,E-E')\times R(E)]$, was then constructed by using it to convolute the 
so-called ``zero target thickness" spectrum to yield the experimental Yale spectrum exactly. The 
first factor, the function $G(E,E-E')$ 
corresponds to the simple folding over the catcher thickness, 
examples of such a folding have been shown above (e.g. Sec. \ref {sec:france}, 
Fig. \ref {fig:n16stopfrance}). Because this 
folding is still inadequate, a further correction factor, $R(E)$, the second factor above, was
required. This multiplicative correction factor $R(E)$ was then determined precisely 
by taking the ratio of the experimental curve to the curve convoluted with $G(E,E-E')$.
Note that in this procedure, 
$G(E,E-E')$ does not have to be realistic, as $R(E)$
takes into account all shortcomings inherent in the process. In fact, $G(E,E-E')$ can
be set to 1.0 at all energies with the same final result. Again a simple 
arithmetic function is used to perform a deconvolution and the response function
is not independent of the actual $\alpha$-spectrum.

The procedure described represents the `semi-empirical' \cite {Zha93}
determination of the response function. 
The circular logic of this data reduction is obvious: 
if the ``folding'' of the ``zero thickness spectrum",
i.e. the spectrum of Ref. \cite {Waf71}, with  $G(E,E-E')\times R(E)$ yields exactly the 
experimental spectrum of Ref. \cite {Zha93}, then the ``unfolding" of the 
spectrum of Ref. \cite {Zha93} with $G(E,E-E')\times R(E)$ will yield exactly the spectrum
of Ref. \cite {Waf71}. Expressed differently: the data of of Ref. \cite {Zha93} are a reproduction
of those of Ref. \cite {Waf71} within the energy region given in Ref. \cite {Waf71},
albeit with larger error bars\footnote {There is no description
how the errors were propagated, but it appears, they also
went from original to final by the same division.} 
and fewer total counts. This explains the major part of the low
$\chi^2$/point=0.17 in the total $R$-matrix fit to these data. 

It may be argued that the 
identity derived above was intended and that the procedure was designed 
specifically to provide a means of extrapolating the correction factor $R(E)$ down
to the low energy region, and that data points above E$_{\alpha}$=1500 keV are thus
irrelevant to the analysis. While such a method would remain
unphysical, there is no evidence anywhere in Refs. \cite {Zha93,Zha93b}
to indicate such a possibility. Factually in any serious $R$-matrix
analysis, these points at higher energy are of utmost importance. 
As displayed in their Fig. 1b, the $R$-matrix analysis
in Ref. \cite {Zha93} suggests indeed the use of the entire spectrum.

\begin{figure}
\center
\includegraphics[width=8.5cm]{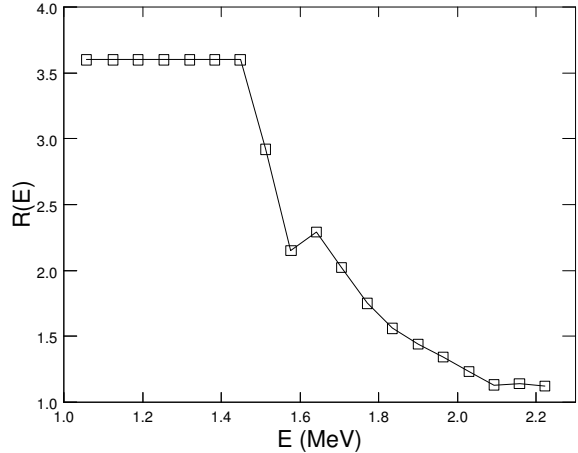}
\caption{The function $R(E)$ as derived in Ref. \cite {Zha93} versus
center-of-mass energies for the lower energy part of the spectrum.}
\label {fig:zhaores}
\end {figure}

The extrapolation of $R(E)$ to lower energies is not described in Ref. \cite {Zha93}; in
Ref. \cite {Zha93b} the following quotation can be found (Sec. 4.1, p.87):
\begin {quotation}
`Consequently, the unfolding procedure carries small uncertainty down to 1.5 MeV
and we further extrapolate it down to 1.1 MeV.  Since $R(E)$ reflects a tail from 
higher energies, the extrapolation is such that $R(E)$ below 1.45 MeV is a 
constant value of 3.6.'
\end {quotation}
With this statement the function $R(E)$ for the low energy part of the
spectrum is shown in Fig. \ref {fig:zhaores}.
It is clear that this choice of an arbitrarily constant function $R(E)$ after a steep 
increase in the ``known" region is unjustified, even, if the deconvolution procedure
was correct.
More reasonably $R(E)$ should follow some functional form 
which is essentially {\it unknown} below E=1.45 MeV; and could, in principle, even increase
or decrease rapidly with increasing energy.
However, $R(E)$ is not a response function of the system by any conventional
use of the word \footnote {A response function is defined by the
energy dependent function (spectrum) derived from the propagation of 
events from a monoenergetic ($\alpha$) source through the experimental
apparatus.}, but an arithmetic divider relying on {\it preconceived}
knowledge. Thus $R(E)$ has no experimental basis.

Applying the constant
$R(E)$ to the initial spectrum shows immediately that the low energy points are still above
the final points, so that some different kind of `deconvolution' step must have been involved. 
A suggestion of what was really done is given in the following quotation
from Ref. \cite {Zha93b} (p.87):
\begin {quotation}
`In fig. 43 we show the convolution of the theoretical curve corresponding
to S$_{E1}$=95 keV-barn with the response function, together with the 
experimental data.'
\end {quotation}

Thus, the $\beta$-delayed $\alpha$-decay spectrum of $^{16}$N as
presented in Ref. \cite {Zha93} is, in the high energy region, a close reproduction 
of the data of Ref. \cite {Waf71}, while for the low energy points it closely follows
a purely theoretical curve.
It does not come as
any surprise that the final result of Ref. \cite {Zha93} is indeed 
S$_{E1}$(300)= 95 keV b.

It is also noteworthy that the 
fractional errors of the low energy data points of Fig. 1(b) \cite {Zha93} are no larger
than those of Fig. 1(a) in spite of the major unfolding procedure required. 
If Fig. 1(b) is converted to counts, some of the lowest energy points have about 
20 counts/channel for which statistical errors of {\it at least} $\pm$25-30\%
are expected, even excluding errors from the unfolding procedure. However,
e.g., for the energy point at 1.448 MeV, the yield is 20.2$\pm$2.5, i.e. with an error 0.56 of
the expected square root of the number of counts. If realistic errors
were taken into account, the errors on $R(E)$ should be added in quadrature
with those resulting from statistics, when the data reduction is done. 


There has been a suggestion that the errors in the 
$\beta$-delayed $\alpha$-decay spectrum of $^{16}$N of Ref. \cite {Zha93}
are correlated. It is, however, abundantly clear from the above procedure
that the final errors should be uncorrelated, if the initial experimental errors 
are uncorrelated. The unfolding procedure
is achieved simply by division of correction factors, a procedure that does not induce 
correlations between data points. There has, however, been a claim in Ref. \cite {Zha93b}
that the errors of the data points are correlated as the detector resolution
is about 50 keV, while the catcher thickness is up to 250 keV. Such a claim
would be incorrect, unless some procedure had been employed which moves events
among the channels {\it after} the data acquisition. There is no evidence of such 
data shuffling anywhere in Refs. \cite {Zha93,Zha93b} as each 
point in the spectrum is divided by an individual number labeled as the
response correction. The low $\chi^2$/point found in
the $R$-matrix fits is completely explained by the fact that the 
$\beta$-delayed $\alpha$-decay spectrum of $^{16}$N of Ref. \cite {Zha93} is basically
a reproduction
of the high-statistics spectrum of Ref. \cite {Waf71} over most of the energy range,
combined with a normalization to a theoretical curve for the lower energy data.

\subsection {The Seattle spectrum}

In 1993/94 a measurement of the $\beta$-delayed $\alpha$-spectrum of $^{16}$N was
started at the University of Washington in Seattle. While this measurement was never 
published by those who carried out the measurement, 
the data were, however, distributed for comparison and 
made available for publication by other groups. 
Particularly the spectrum is published in tabular form in
Ref. \cite {Fra07}.
The only, rather incomplete, information available about 
the experiment is from two annual research reports of 1994 \cite {Ade94}
and 1995 \cite {Ade95}.

The experiment used a Ti$^{15}$N target bombarded by a deuterium beam
to produce $^{16}$N which was partially captured in 
carbon foils of 10 or 20 $\mu$g/cm$^2$ thickness. The activity was then transfered
to a position between two silicon detectors, and $\alpha$ and $^{12}$C particles were 
observed in coincidence,
as in Ref. \cite {Azu94}. It is not clear whether the distributed spectrum was made with 
a 10 or 20 $\mu$g/cm$^2$ thickness, or, if both spectra were combined.
Other unknown information includes the method of energy calibration 
and the resolution of the
detectors. It is, however, of interest to note the following remark in Ref. 
\cite {Ade94}:
\begin {quotation}
`In Fig. 1.2-1, we show the two dimensional histogram of carbon energy 
versus alpha-particle energy. The contribution of the broad 1$^-$ state 
at 9.6 MeV in $^{16}$O is the dominant feature
of the histogram, while the true low energy events are visible in a {\it curved}\footnote
{Emphasized here.} group roughly along the diagonal line.'
\end {quotation}
Certainly, in the measurements of Refs. \cite {Azu94,Tan07} which used approximately 10~$\mu$g
carbon collector foils, the pulse height relation between $^{12}$C and $\alpha$ events was
strictly linear. The curvature may suggest that more mass per unit area than stipulated 
in Ref. \cite {Ade94}, possibly caused
by carbon deposition during bombardment, was present in the collector foil, 
or that the data acquisition
had some other problem. It is not clear what measures, if any, were taken
to prevent carbon build-up on the foils, or if the foil
thickness was otherwise monitored. In any case, accurate 
energy calibration of a curved group in the
two dimensional spectrum appears difficult. The method of energy calibration was,
however, not reported (see Sec. \ref {sec:compsea}). No coincidence efficiency
measurement is described.

The actual geometry, details such as beam spot size and detector distances, 
of the Seattle experiment are not known. Therefore only a simulation
using very generic features can be performed, see Sec. \ref {sec:seattlesim}.

\subsection {The Argonne measurement}

Recently a paper was published \cite {Tan07} in which a measurement of the 
$\beta$-delayed $\alpha$-spectrum of $^{16}$N at Argonne National Laboratory
is described. In this experiment, a fast $^{16}$N beam produced
by the d($^{15}$N,p)$^{16}$N reaction was slowed down in a
gas cell and captured onto a target carbon foil with a claimed
beam spot of 5 mm. Within 15 s, the collected
activity was rotated into an ionization chamber where both the $\alpha$-particle
and the $^{12}$C recoil were detected. A similar ratio cut as in Ref. \cite {Azu94}
has been applied, eliminating degraded target and detector events.
Because of the particular geometry, events hitting the target frame were
also visible in the two dimensional pulse height spectra and were eliminated.
220,000 events were published. The energy calibration used the $^{10}$B(n,$\alpha$)$^7$Li
and the $^6$Li(n,$\alpha$)t reactions, with the lowest energy calibration line
at 1.472 MeV $\alpha$ energy. The electronic coincidence efficiency was checked with
a pulser; however, no check of the coincidence efficiency for the full system
has been presented. The energy resolution is quoted with 40 keV at 1.4 MeV;
no response function of the detector is reported.


\section {GEANT4 simulations of $\beta$-delayed $\alpha$-decay spectra of $^{16}$N}
\label {sec:geant}

\subsection {Introduction}


For extended Monte-Carlo (MC) simulations, the GEANT4 \cite{geant4} toolkit has  been used. 
It provides excellent possibilities for the definition of complex geometries and 
particle tracking therein, but lacks in support  for processes often relevant 
for nuclear physics. In particular, there is no process for Coulomb scattering of 
ions at low energies, and the  energy loss tables are not up-to-date. The built-in multiple 
scattering process is only useful for high-energy light particles  scattered on 
heavy ions, neglecting the energy transfered to the recoil particle. 
Therefore, a custom-made Coulomb scattering process has been developed and added to GEANT4
to test primarily ion propagation in ionization chambers. This process
has been applied here.

GEANT4 asks in each process for the mean free path, and takes the smallest
of these values, or the distance to the next geometrical boundary,  
applies all "continuous" processes (only the energy loss in our case), 
and finally all "discrete" processes (only Coulomb scattering in our case).  
The custom-made Coulomb scattering process as well as the  energy loss process 
are described in the following.

\subsubsection{The Coulomb scattering process}
The code is based on a paper from M\"oller, Pospiech, and Schrieder  
\cite{moeller}, describing the implementation of a Monte Carlo (MC) simulation for a chain of 
single scatterings on a screened Thomas-Fermi potential.  
Using reduced energy and reduced cross-section quantities the problem can be simplified 
to the integration of one scalar function (see. Ref. \cite{lindhard} for the tabulated function) 
for all charges, masses,  energies, and angles. In this approach, the total cross section 
is a free parameter; the larger the cross-section, the more scattering will be calculated, 
and the result will be more accurate for small  scattering angles. The physical limit is 
the half-distance $r_0$  between the atoms of the material the particle is moving in, 
$r_0 =  0.5 \,\, n^{-1/3}$, where $n$ is the atomic density of the material.  
A lower limit is given by the screening radius 
$a = 0.8853 \,\, a_B / \sqrt{Z_1^{2/3} + Z_2^{2/3}}$, with $a_B$ as Bohr radius. 
The reduced  total cross-section $J_{tot}$ is defined as $J_{tot} = \sigma / \pi  a2$. 
For the calculations here we used values of $J_{tot}$ = 1-10.

For each scattering, the change of direction and energy is calculated  and the 
recoil ion is generated if is above a threshold (typically 1 keV). 
GEANT4 also tracks the recoils, so they obey the same processes 
as defined for the primary ions.

\subsubsection{The energy loss process}

The stopping power calculation in GEANT4 is based on the Ziegler 
parameterization \cite{ziegler} for stopping of protons in all matter, 
and on the effective charge model to scale the energy loss  from protons 
to higher charged ions. However, different sets (tables)  of coefficients 
can be selected based on different evaluations. The  default set is based on 
the ICRU 49 report \cite{icru} but available  are also tables from Ziegler 1977, 
Ziegler 1985, and SRIM 2000. For  the ICRU and Ziegler tables two variants 
can be selected, based on  helium stopping powers and on 
proton stopping powers. For the  calculations here, we used the ICRU-49 table based 
on He stoppings but  we also compared the results with calculations 
based on other tables.  The differences are not significant.

\subsection {The Mainz spectrum}
\label {sec:mainzsim}

\begin{figure}
\center
\hspace*{-0.5cm}
\includegraphics[width=6.5cm,angle=0]{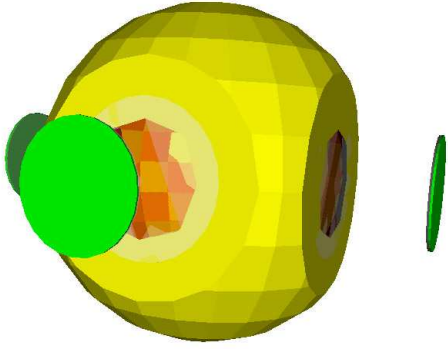}
\caption{(color online) GEANT4 geometry used to simulate the Mainz experiment \cite {Neu74}.
         The figure shows the gas cell and the outside detectors. $\alpha$-particles
         and $^{12}$C nuclei are multiple scattered by the gas, the collodion foil, and
         in the gold layer of the detector surface.}
\label {fig:geantgeowaef}
\end {figure}

The exemplary description of the Mainz experiment presented in Ref. \cite {Neu74}
allows one to simulate this measurement with a high degree of confidence. The 
geometry used in the GEANT4 simulation is shown in Fig. \ref {fig:geantgeowaef}.
The  fiducial volume, i.e. the volume from where $\alpha$ 
particles can reach the detectors is identical to the one of
Ref. \cite {Neu74}.

\begin{figure}
\center
\hspace*{-1.0cm}
\includegraphics[width=6.5cm,angle=-90]{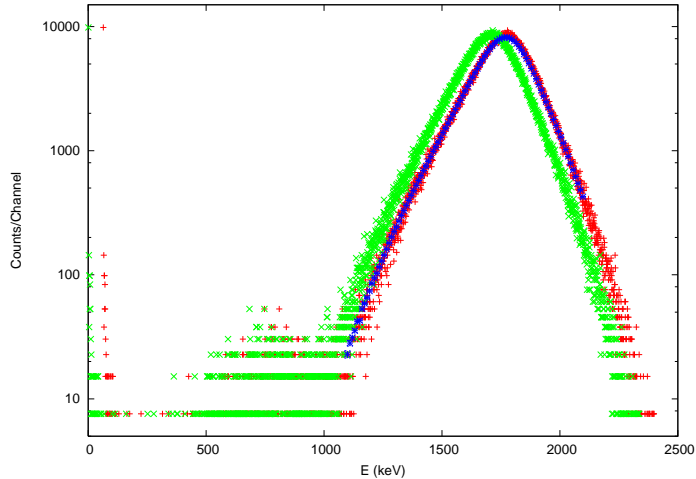}
\caption{(color online) GEANT4 simulations and comparison with the Mainz spectrum. The
 simulated spectrum is shown with the x-points, the simulated, shifted 
 (63 keV) spectrum is shown by the cross points (+), and the experimental Mainz
 spectrum by the star (*) symbol. The latter two spectra
 overlap almost perfectly. The energy is in the laboratory system. }
\label {fig:geantwaefsim}
\end {figure}

The results of the simulation are shown in Fig. \ref {fig:geantwaefsim}.
The simulations (using the TRIUMF initial distribution) show that the
Mainz spectrum would be shifted downward by about 60 keV, if an intrinsic
calibration had not been applied. Applying such a shift shows that
the experimental and the simulated Mainz spectrum show excellent agreement.
The shift and broadening relative to the TRIUMF spectrum confirms
that any notion of a `zero mass spectrum' attributed to the Mainz measurement in 
Ref. \cite {Zha93} is not justified.

\subsection {The Yale spectra}

\subsubsection {GEANT4 Simulations of the Yale spectra}
\label {sec:simyale}

\begin{figure}
\center
\hspace*{-0.5cm}
\includegraphics[width=6.5cm,angle=-90]{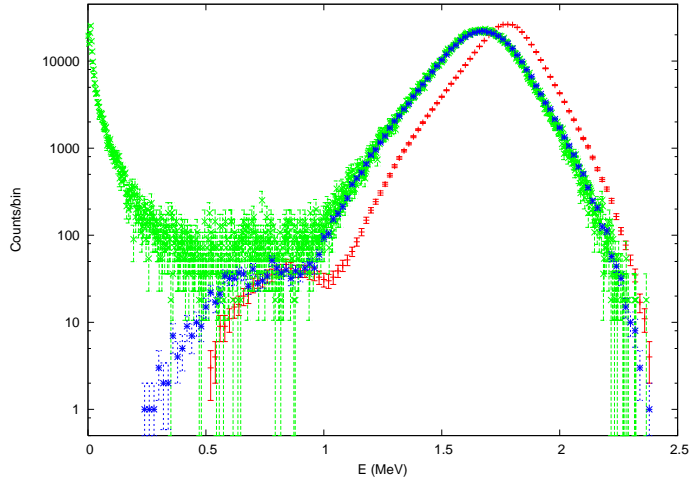}
\caption{(color online) Comparisons of the original spectrum (bars) used in the 
simulation (TRIUMF), a spectrum with energy loss only applied (asterisks), and
one with full scattering (crosses) from GEANT4 for a point source extending down to zero energy.
The energy is for $\alpha$-particles in the laboratory system.}
\label {fig:geants}
\end {figure}

The information available about the Yale experiments was used to simulate
them. A $^{16}$N source was uniformly
distributed through a 180$\mu$g/cm$^2$ aluminum foil, at first without any
extension in area. All angles $\theta$ were
simulated, where $\theta$ is the initial emission angle of the $\alpha$
particle relative
to the foil. (Zero degrees is perpendicular to the foil.)
The spectrum derived from the TRIUMF R-matrix fit \cite {Azu94} 
was used in most studies for the original distribution, excluding thus the
detector resolution in the measured TRIUMF spectrum. Also the
final spectrum of Ref. \cite {Fra07} was employed 
once. The result of our kind of simulation 
is equivalent to a convolution of the initial spectrum and always
leads to a `broader' spectrum. Thus it is pointless to use as an initial spectrum one which
is already `broader' or of the same width, as no agreement will result. 
Fig. \ref {fig:geants} shows a 
comparison between the original input spectrum, the spectrum subjected to 
a thick catcher as in Fig. \ref {fig:n16stopfrance} and the GEANT4 calculations
for a detector opening angle of 10$^{\circ}$.
It is obvious that the GEANT4 simulated spectrum agrees very well with the 
previously simulated
spectrum (Fig. \ref {fig:n16stopfrance}) 
that includes only catcher thickness effects. However, at the low energy side,
a tail develops caused by small to large angle scattering obscuring the physical low
energy plateau of the $^{16}$N spectrum and eventually leading to 
many very low energy events.
Those would typically be below the detection threshold, or obscured by $\beta$ rays from
the abundant $^{16}$N decays.

\begin{figure}
\center
\hspace*{-0.5cm}
\includegraphics[width=6.5cm,angle=-90]{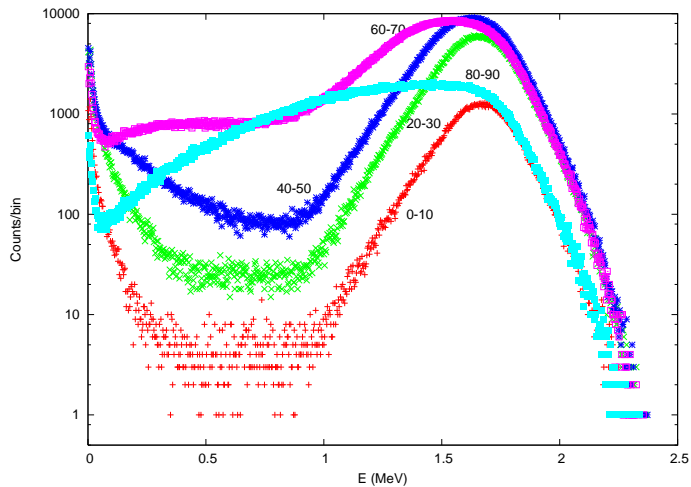}
\caption{(color online) The GEANT4 simulated spectrum for different angular ranges
of $\theta$, as indicated by the labels for the Yale conditions.}
\label {fig:geantangle}
\end {figure}

The spectrum deteriorates with increasing opening angle of the detectors. This is
shown in Fig. \ref {fig:geantangle} displaying different angle ranges.
As can be seen, there is a target thickness effect by projection, making the
main peak wider as $\theta$ increases. In addition, the Coulomb scattering becomes more and
more significant, until at high angles, it obscures the
underlying spectrum completely. The total counts in the spectra of Fig. 
\ref {fig:geantangle} scale with 
$\sin{\theta}$ until at angles close to 90$^{\circ}$ the foil itself
obscures detection.

\begin{figure}
\center
\hspace*{-0.5cm}
\includegraphics[width=6.5cm,angle=-90]{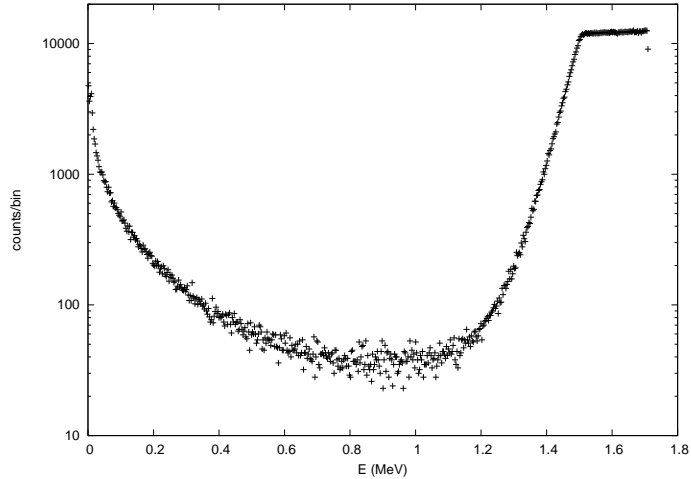}
\caption{Response function using the Yale setup for a monoenergetic
$\alpha$-source at E$_{\alpha}$=1.71 MeV showing the energy loss
range (plateau), a short range tail (down to about 1.2 MeV),  and a long range tail
with a many very low energy events. The energy is in the laboratory system.}
\label {fig:geantresp}
\end {figure}

In a subsequent step, we have simulated the geometry of the France III experiment
\cite {Fra96,Fra07}, and, as the geometry is similar, the
Zhao experiment\cite {Zha93}, including the extended catcher foil and the large-area silicon
detector array. We have derived both single energy response functions for each of
the energy points in the France III spectrum, as well as an integrated spectrum from
the TRIUMF distribution.
Fig. \ref {fig:geantresp} shows the single energy response function for
E$_{\alpha}$=1.71 MeV, close to the main peak of the spectrum.
As is visible, the response function shows the typical broadening by
the catcher foil energy loss (plateau) with a short range tail going down
to about 1.2 MeV, followed by a long range tail leading to many very low energy
degraded events.

\begin{figure}
\center
\hspace*{-0.5cm}
\includegraphics[width=6.5cm,angle=-90]{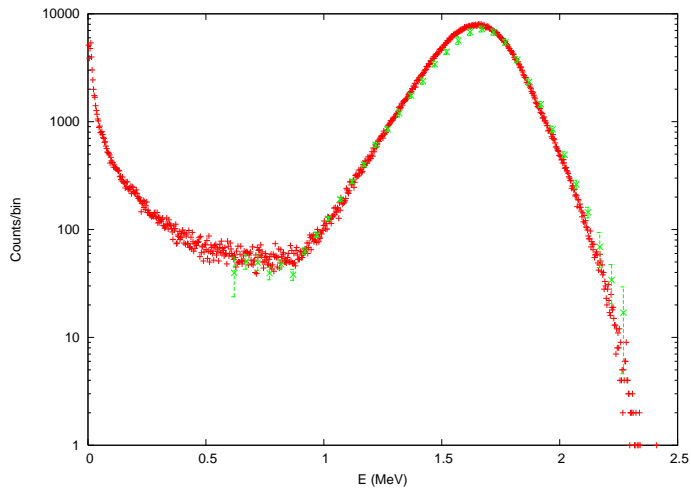}
\caption{(color online) GEANT4 simulation using the Yale set-up for an 
$\alpha$-spectrum corresponding to the TRIUMF distribution.}
\label {fig:geantfrance-triumf}
\end {figure}

In Fig. \ref {fig:geantfrance-triumf} a simulation is shown which uses
the geometry of the Yale experiments and the TRIUMF $\alpha$-particle 
energy distribution.
No detector energy resolution has been included into the simulations which 
may partially explain the slightly wider than simulated France III spectrum 
in the high energy region. 
We also suggest that the $\beta$-efficiency correction in the
high energy region of Ref. \cite {Fra07} is somewhat overdone 
(see Fig. \ref {fig:france-nobeta}). The simulation
tail is slightly higher than the measured one. This might be corrected by assuming
a nonisotropic or nonhomogeneous distribution of $^{16}$N in the catcher
foil, but there is no information about such a possible distribution.

\begin{figure}
\center
\hspace*{-0.5cm}
\includegraphics[width=6.5cm,angle=-90]{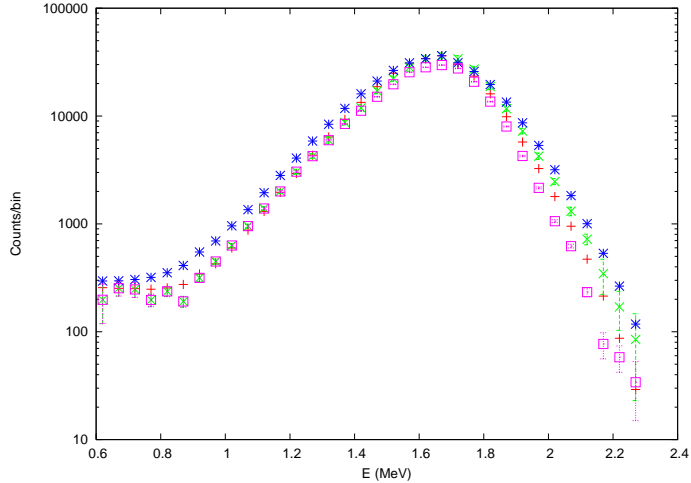}
\caption{(color online) 
Spectral response by response functions taken at single energy points 
(see Fig. \ref {fig:geantresp}) employing the Yale set-up for an 
$\alpha$-spectrum corresponding to the TRIUMF spectrum (bar) and the final France III spectrum 
(asterisks) in comparison with the France III $\beta$-corrected (cross) and 
$\beta$-uncorrected spectra (square). The simulation spectra have been folded with a 55 keV
Gaussian detector response \cite {Fra96}.}
\label {fig:geant-ff-tr}
\end {figure}

As discussed above, response functions for each energy in the France III spectrum were
calculated. From these similar spectra to that shown in Fig. \ref {fig:geantfrance-triumf}
can be derived for both the TRIUMF and the France III final spectrum.
These are shown in Fig. \ref {fig:geant-ff-tr} in comparison with the France III raw spectrum
and the non-$\beta$-efficiency corrected spectrum (see also Fig. \ref {fig:france-nobeta}).
As in Fig. \ref {fig:geantfrance-triumf}, Fig. \ref {fig:geant-ff-tr} shows the
simulated spectrum with the TRIUMF initial spectrum to be between the $\beta$-corrected and 
$\beta$-uncorrected spectra in the high energy region, 
while the France III spectral input leads to points
above the corrected spectrum. However, the agreement with the France III spectrum
at the low energy side, where $\beta$-efficiency corrections are negligible, is excellent
for the case of the TRIUMF spectral input, while for the France III final spectrum
as input there is a considerable
discrepancy. The simulations
also demonstrate that the six lowest energy points in the France III raw spectrum are not
related to the plateau in the spectrum at low energies found at TRIUMF, but to the 
tail in the response function mainly caused by Coulomb scattering in the thick catcher
foil. 

In general, the simulations show that the measurements at TRIUMF and Yale
are in reasonable agreement (for both France III and Zhao), 
while the final France III spectrum is in disagreement with
the original spectrum. 

In the subsequent section, we investigate, by applying a deconvolution 
method described in literature,
the degree to which useful information can be extracted from the Yale measurements.

\subsubsection {Deconvolution of the Yale spectra}

\label {sec:yaledecon}

Deconvolution of experimental information is, in general, a difficult
problem. There are several techniques known, all
of them iterative, but not convergent. 
In the case of the $\beta$-delayed $\alpha$-decay of $^{16}$N, 
we are dealing with a spectrum, i.e. a
finite number of discrete, positive data points, something we
also expect as the final solution. For such a problem the Gold algorithm
of deconvolution is most appropriate \cite {Mor05}\footnote {In their introduction
the authors of Ref. \cite {Mor05} make the following statement about deconvolution:
`From the numerical point of view, deconvolution belongs to one of the 
most critical problems. It is a so-called ill-posed problem, which
means that many different functions solve the convolution equation within the
allowed error bounds. The estimates of the solution are extremely sensitive to errors
in the input data. Very frequently the noise present in the 
input data causes enormous oscillations
in the results after deconvolution.'}. The data analysis frame 
`ROOT' \cite {Roo07} developed at CERN 
provides an implementation of the Gold deconvolution algorithm that
we have applied here, in particular by the TSpectrum::Unfolding()
function. For the response function, necessary to be
known in the deconvolution, the GEANT 4 simulated
spectra for each of the raw Yale data points, as discussed
in Sec. \ref {sec:simyale}, Fig. \ref {fig:geantresp}, are employed. 

\begin{figure}
\center
\hspace*{-0.5cm}
\includegraphics[width=9.5cm]{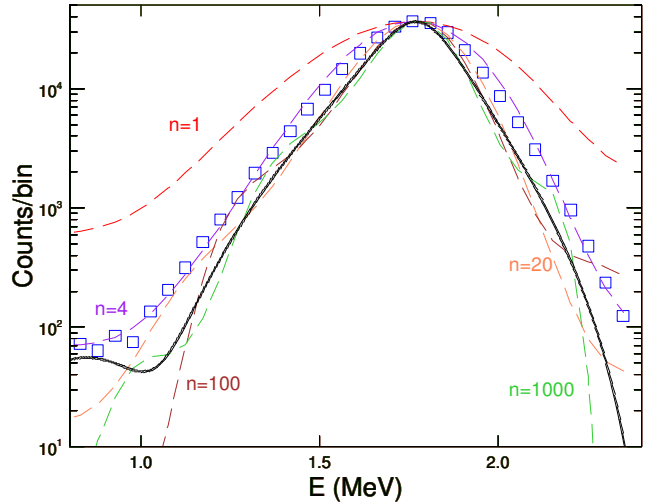}
\caption{(color online) Deconvoluted spectra
according to the Gold algorithm for increasing numbers of iteration steps (n=1-1000)
as indicated by the labels on the dashed lines.
The France III final spectrum is shown as the boxes and the TRIUMF $R$-matrix 
fit as straight line. The energy is in the laboratory system.}
\label {fig:yaledecon}
\end {figure}

As the deconvolution in the Gold algorithm is an 
iterative process, the number of iterations itself is a free parameter
to be chosen by judgment as the process, in general, is not very
well convergent. For the initial iterations the
spectrum gets narrower and narrower while the low energy tail gets smaller.
We find only a mild convergence for the high statistics points
at the maximum of the $\beta$-delayed $\alpha$ decay distribution of $^{16}$N
for an increasing number of iterations starting at about 20 iterations,
while the low count regions at both the low and the high energy 
sides of the $^{16}$N spectrum start to fluctuate with an increasing number
of iterations. The width of the main peak converges to the
TRIUMF width or to a value slightly below. In Fig. \ref {fig:yaledecon} deconvoluted spectra
obtained by the Gold algorithm for increasing iteration steps are shown.

From this deconvolution process applied here it can be 
concluded that, without any pre-conceived knowledge
of the shape of the $\beta$-delayed $\alpha$ spectrum of $^{16}$N, only the
presence of the major peak with some reasonable width can be concluded, while any other
information is obscured by the experimental procedure chosen in the
Yale measurements. Poor energy resolution cannot be reversed.

\subsection {Simulations of the Seattle and TRIUMF experiments} 

\label {sec:seattlesim}

\begin{figure}
\center
\hspace*{-0.5cm}
\includegraphics[width=6.5cm,angle=-90]{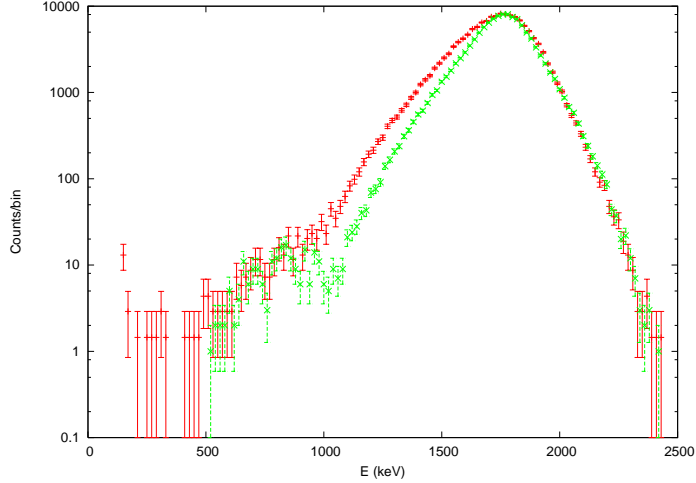}
\caption{(color online) Comparison of a simulated spectrum using a 10 $\mu$g/cm$^2$ C foil (x)
and a 100 $\mu$g/cm$^2$ C foil (bar). The energy is in the laboratory system.}
\label {fig:sim10-100cut}
\end {figure}

Using the available information, we attempted to simulate the Seattle
experiment. As this spectrum has used $^{12}$C+$\alpha$
coincidences it was first simulated, how, despite the ratio cut
\footnote {We call a `ratio cut' a condition on the spectra where
only a limited range of ratios of apparent $\alpha$ and $^{12}$C 
events is considered to be valid. If there were no resolution
and pulse height issues, the energetic ratio between $\alpha$-
and $^{12}$C events would be exactly 3. For a description and application of such
a condition, see Ref. \cite {Azu94}.},
the spectral form changes with increasing target foil thickness.
What happens for a 10$\mu$g/cm$^2$ C foil  
in comparison to a 100 $\mu$g/cm$^2$ C foil is shown in Fig. 
\ref {fig:sim10-100cut}. The spectrum of the 100 $\mu$g/cm$^2$ C foil
has been shifted higher in energy to match that of the 10 $\mu$g/cm$^2$
C foil. In both simulations, the spectra use the TRIUMF input.
It is clear that while tail events stay relatively well preserved, there is 
a short range tail component that is not easily removed by the ratio
cut. Therefore also the method of coincidence ratio cuts 
calls for as thin foils as possible.

\begin{figure}
\center
\hspace*{-0.5cm}
\includegraphics[width=6.5cm,angle=-90]{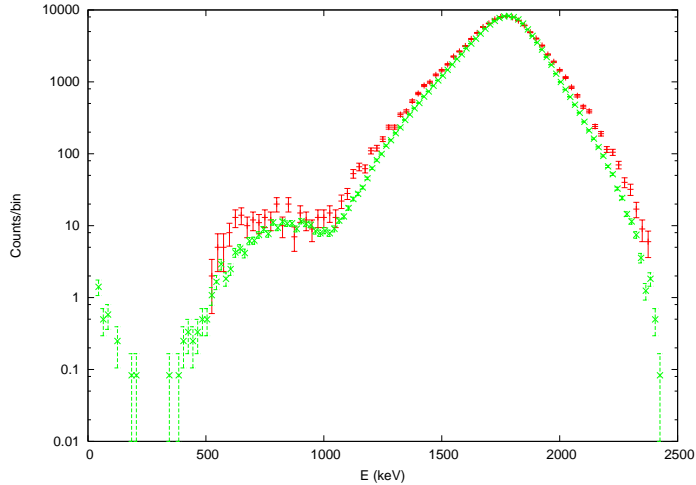}
\caption{(color online) Comparison of a simulated spectrum using a 20$\mu$g/cm$^2$ C foil (cross)
and the Seattle spectrum (bar). The energy is in the laboratory system.}
\label {fig:comp-sea-20}
\end {figure}

In Fig. \ref {fig:comp-sea-20} a comparison between the TRIUMF
based simulated spectrum based on a 20$\mu$g/cm$^2$ C foil and the Seattle
spectrum is shown. 
A slight energy adjustment, corresponding to the energy
loss in the foil has been done.
Both the high and the low energy side of the Seattle
spectrum are wider than in this comparison, see also the discussion in Sec. 
\ref {sec:compsea} 
regarding energy calibration. 
It may be noted that approximate agreement
on the low energy side can be obtained with a foil thickness
of approximately 50 $\mu$g/cm$^2$. However, no possible foil thickness
in the simulation can reproduce the high energy side of the 
spectrum from the TRIUMF input.

As seen for the simulations to the Seattle data a 10 $\mu$g/cm$^2$ C foil introduces
little disturbance to the $\beta$-delayed $\alpha$ spectrum of $^{16}$N. We have
used the TRIUMF geometry and input to simulate the TRIUMF data as well. The agreement is
good after folding with the approximately 30 keV energy resolution of the detectors.

\subsection {Simulations of the Argonne experiment}
\label {sec:argonnesim}

\begin{figure}
\center
\hspace*{-0.5cm}
\includegraphics[width=8.5cm,angle=0]{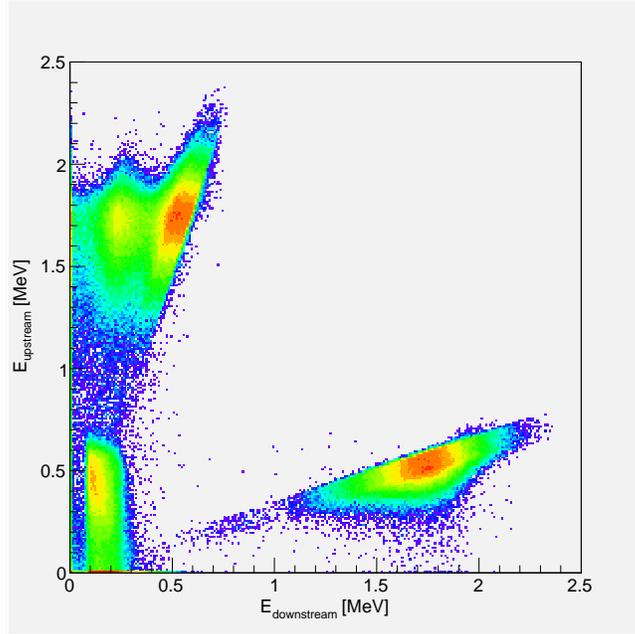}
\caption{(color online) Simulated two dimensional energy versus energy 
spectrum for a gas chamber with 150 Torr as described for the
Argonne experiment. The asymmetry in regard to the 45$^{\circ}$ axis 
originates from the asymmetric collector foil mounting in reference to the foil
supporting frame. The energies are in the laboratory frame.}
\label {fig:simar2d}
\end {figure}

Following the description of the experiment \cite {Tan07}, the spectra have been
simulated in GEANT4 based on the TRIUMF spectrum as previously. No $^{16}$N isotopes
in gaseous form and no noise, as reported in Ref. \cite {Tan07},
have been included. The two-dimensional energy versus energy spectrum
from the simulations is shown in Fig. \ref {fig:simar2d}.
Particles were tracked through the P10 gas (90\% argon, 10\% CH$_4$)
to complete stop either in the gas or a boundary. 
The reported partial energy collection
for high energy events for ion chamber pressures of
150 Torr \cite {Tan07} are confirmed.

\begin{figure}
\center
\hspace*{-0.5cm}
\includegraphics[width=6.5cm,angle=-90]{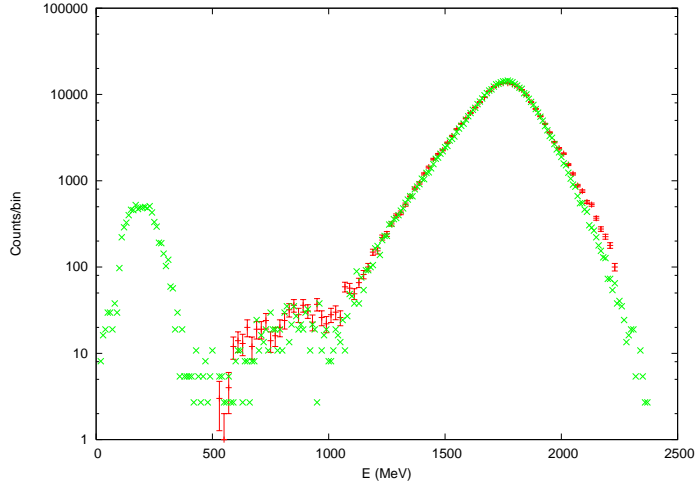}
\caption{(color online) Comparison between the simulated (bar) and the Argonne
spectrum (x). A pressure of 195 Torr is used. The energy is the 
$\alpha$-energy in the laboratory in keV.}
\label {fig:simarcomp}
\end {figure}

In Fig. \ref {fig:simarcomp} the comparison between the simulated 
GEANT4 spectrum and the spectrum of the Argonne work is shown. In the simulated
spectrum a condition (cut) for maximum carbon energy (channel) as well as a
ratio condition were applied. 
As the numerical value for the latter is not reported in 
Ref. \cite {Tan07}, we applied one similar to Ref. \cite {Azu94}.
The figure shows reasonable agreement on the low energy side of the
main peak, with a slight excess of counts in the simulated spectrum
around E$_{\alpha}$=1500 keV
and E$_{\alpha}$= 1 MeV, see also the discussion in Sec. \ref {sec:comptarg}. 
The high energy side of the spectrum is definitely wider, but no detector resolution
effects have been included in the simulation. The simulation 
demonstrates also that measurements using ionization chambers have 
potential problems with the response tail from the target, as this limit seems
to be reached in the Argonne measurement. In conclusion: the Argonne measurement
is consistent with the TRIUMF one at least for the region of 
E$_{\alpha}$=1.1 to 1.3 MeV. For further discussion, see Sec. \ref {sec:comptarg}.


\section {Comparison and R-matrix fits to different data}
\label {sec:compmeas}

\subsection {Fits and comparison of spectra}

Independent of their credibility all of the $\beta$-delayed $\alpha$ spectra of $^{16}$N 
can be subjected to $R$ matrix fits. In particular, this can show how much the
apparent differences between the spectra influences the deduced S-factor S$_{E1}$(300).
Only fits with an interaction radius of $a$=5.5 fm and with the $\beta$
feeding factors $A_{23}$=0 and 
$A_{33}$=0 for the $f$-wave (as in Ref. \cite {Azu94}; also see Ref. \cite {Azu94}
for $R$-matrix notation) 
were performed, as it is not the goal
of this paper to produce a final combined result with a variation 
of all parameters. However, it must be stressed again that the
results derived from meaningless spectra are meaningless.

Comparisons of the final experimental spectra are as long meaningless
as all of the measurements have different energy resolutions and different
$^{16}$N source thicknesses. In principle,
only unconvoluted $R$-matrix fits, with convolutions applied in the fittings,
should be compared, as clearly a spectrum with a detector energy resolution of
10 keV will be different from one with a resolution of of 100 keV, even, if the detector
responses would be simply Gaussian over many orders of magnitude.

\subsubsection {Comparison between the France III and the Zhao spectra}
\label {sec:compyale}

\begin{figure}
\center
\hspace*{-0.5cm}
\includegraphics[width=6.5cm,angle=-90]{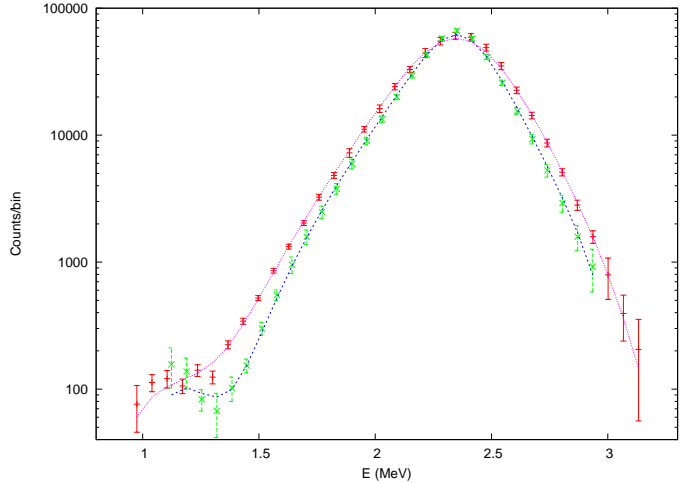}
\caption{(color online) Comparison of the final $\beta$-delayed $\alpha$-spectra of 
Ref. \cite {Fra96,Fra97,Fra07} (bars) and Ref. \cite {Zha93} (crosses) with both
spectra normalized in height to the one of Ref. \cite {Azu94}. The
energy axis is the center-of-mass-system.
Also shown are $R$-matrix fits to these spectra.}
\label {fig:zha-france}
\end {figure}

In Refs. \cite {Fra96,Fra97,Fra07} no comparison between the $^{16}$N 
spectrum presented there and the one of Ref. \cite {Zha93} is given, even though
differences with the spectrum of Ref. \cite {Azu94} are pointed out with
considerable detail.
Both spectra \cite {Fra07,Zha93}, together with R-matrix fits to them, are shown in
Fig. \ref {fig:zha-france}.
Several observations may be made: the two spectra do not agree, not only
on the low energy side of the main peak, but also on the high energy 
side even though they both originate from the
same initial distribution. The quality of the $R$-matrix fits
for both spectra is more than excellent, as 
already noted above for the case of Ref. \cite {Zha93}. Because both spectra
have gone through extensive, non-justified, data reduction, it is also
unclear, how the detector resolutions should be treated. E.g. for the
Mainz spectrum \cite {Neu74}, a detector resolution of 12-18 keV (FWHM) at 1.5 MeV
is quoted, with the full system
response leading to about 40 keV at 1.2 MeV $\alpha$ energy,
while for Ref. \cite {Zha93b} 50 keV is given at an
unspecified energy. However, as discussed above, the spectrum of Ref. \cite
{Zha93} has been normalized to that of the Mainz group. Indeed, a low 
energy resolution (20 keV or less) fits the spectrum of Ref. \cite {Zha93} best,
particularly around the maximum. On the other hand, the broad spectrum
of Refs. \cite {Fra96,Fra97,Fra07} is well fitted with the quoted experimental
detector resolution of 55 keV. The particular fits shown in 
Fig. \ref {fig:zha-france} correspond to
S$_{E1}$(300)=79.6 keV b for Refs. \cite {Fra96,Fra97,Fra07} for a bad
resolution fit and S$_{E1}$(300)=101 keV b for Ref. \cite {Zha93} for a 
low resolution fit. Ref. \cite {Zha93} suggests S$_{E1}$(300)=95 keV b,
see also Sec. \ref {sec:diszhao}. No value of  S$_{E1}$(300) has been suggested for
the France III \cite {Fra07} measurement.

\subsubsection {Comparison between the Seattle and the TRIUMF spectra}
\label {sec:compsea}

\begin{figure}
\center
\hspace*{-0.5cm}
\includegraphics[width=6.5cm,angle=-90]{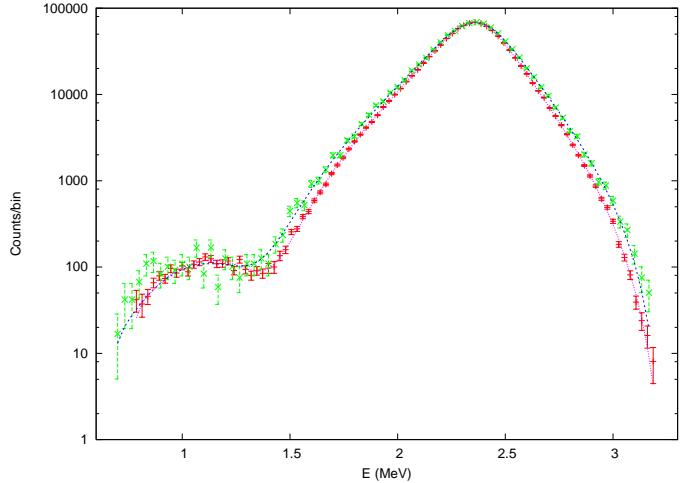}
\caption{(color online) Comparison of the final $\beta$-delayed $\alpha$-spectra of 
Ref. \cite {Ade94} (crosses) and Ref. \cite {Azu94} (bars) with the Seattle
spectrum normalized to that of Ref. \cite {Azu94} at the peak. Energies
are in the center-of-mass system.
Also shown are $R$-matrix fits to these spectra.}
\label {fig:sea-triumf}
\end {figure}

In Fig. \ref {fig:sea-triumf} a comparison between the Seattle \cite {Ade94}
and the TRIUMF spectra are shown \cite {Azu94}.
Obviously, the two spectra disagree, as the Seattle spectrum is
wider both in the low and the high energy regions of the TRIUMF spectrum.
However, as the method of energy calibration for the
Seattle spectrum is unknown to us, it may be noted that a simple linear recalibration
would give a nearly perfect match between the two spectra 
($E_{new}=0.95*E_{old}+0.118$ MeV). Note the following quotation
from Ref. \cite {Ade95}:
\begin {quotation}
`Our results are consistent with the previous measurements at TRIUMF \cite {Azu94}
and Yale \cite {Zha93}\footnote {Citations as quoted here.}'.
\end {quotation}
This suggests that the Seattle group does not consider the differences between those
spectra as important. In fact, the difference between these spectra and the
one of France III is the most pronounced.

The detector resolution in the Seattle experiment is not published. 
In principle, a larger energy resolution can achieve
agreement between the TRIUMF and the Seattle spectra. We have varied 
that resolution and find an optimum value at about 30 keV with 
S$_{E1}$(300)=97.1 keV b. However, fitted simultaneously with the 
radiative $E1$ ground state data, a smaller resolution (20 keV) is preferred with
S$_{E1}$(300)=95.7 keV b. For larger resolutions than 30 keV, S$_{E1}$(300)
declines with a decrease in fit quality. No S$_{E1}$(300) has been suggested by the
Seattle group.

\subsubsection {Comparison of the Mainz and TRIUMF spectra}

A comparison between the Mainz single count spectrum
and the TRIUMF coincidence spectrum is shown
in Fig. 15 of Ref. \cite {Azu94}. It has never been claimed the two both spectra
agree. However, it was found and pointed out in Ref. \cite {Azu94} 
that the TRIUMF singles spectrum,
before application of the coincidence cuts, agrees well with the Mainz spectrum. The difference
with the coincidence spectrum is largely from the coincidence ratio cuts applied to remove
degraded $\alpha$ events found in the coincidence spectrum. In the 
simulations of Sec. \ref {sec:mainzsim} it has indeed be shown that 
the Mainz (single detector) spectrum is fully consistent with the TRIUMF one.

\subsubsection {Comparison between the TRIUMF and Argonne spectrum}
\label {sec:comptarg}

\begin{figure}
\center
\hspace*{-0.5cm}
\includegraphics[width=6.5cm,angle=-90]{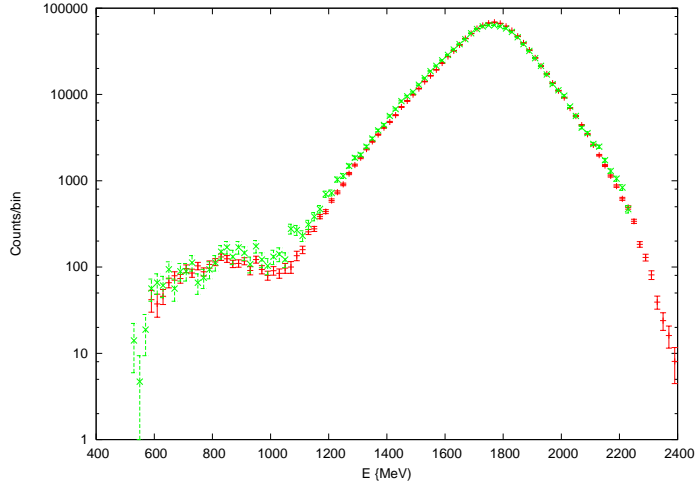}
\caption{(color online) Comparison of the $\beta$-delayed $\alpha$-spectra of $^{16}$N
Ref. \cite {Azu94} (bars) and Ref. \cite {Tan07} (cross).}
\label {fig:argcomp}
\end {figure}

In Ref. \cite {Tan07} a comparison between the low energy part
of the TRIUMF spectrum and the Argonne spectrum as well as 
with the Z. Zhao
spectrum \cite {Zha93}, is presented. 
Better agreement of the Argonne spectrum with the Mainz and 
Z. Zhao spectrum is claimed than with the TRIUMF spectrum.
The comparison is in principle invalid as the TRIUMF combined detector 
and $\beta$-$\nu$ recoil resolution is 30$\pm$5 keV, while
a resolution of 40 keV (no error) at 1472 keV is 
claimed in Ref. \cite {Tan07} ; combined
with the  $\beta$-$\nu$ recoil resolution that averages to 
43 keV\footnote {In the $R$-matrix fits of Ref. \cite {Tan07}
it is not mentioned, if the fits do include a convolution
by the detector resolution. $\beta$-$\nu$ recoil effects
are not mentioned in Ref. \cite {Tan07} and for that
matter in any publication of $^{16}$N spectra, except the 
TRIUMF one \cite {Azu94}.}. Fig. \ref {fig:argcomp} shows a
comparison with the TRIUMF spectrum\footnote{As the Argonne group has
not agreed to make the digitized spectrum publicly available\cite {Reh07}, 
we have read it out from the publication. Minor inconsistencies
particularly for the high count points, are therefore possible.
Errors are
taken as the square root of the counts, as apparently has been
done for the low count points, where errors can be read out.}.

We make the following observations: As described, the region of the Argonne 
spectrum from about
E$_{\alpha}$=1.05 MeV to about 1.3 MeV is above the TRIUMF \cite {Azu94}
spectrum. However, there are other differences: (i) the high energy points
of the Argonne spectrum are above the TRIUMF spectrum starting
at about E$_{\alpha}$=1900 keV. We attribute this to the poorer
detector resolution in the Argonne experiment (see arguments
above for non-compatibility of spectra) and also $^{18}$N background effects. 
(ii) Also in the region near
E$_{\alpha}$=1.5 MeV, the Argonne data are above the TRIUMF data.
In that region, the continuation of the response function
that cannot be removed by the ratio cut in the TRIUMF
spectrum, has been subtracted, see Fig. 10 of Ref. \cite {Azu94}.
No removal of a similar response function in the Argonne spectrum,
nor of their additional background from $^{16}$N in
the gas phase have been reported. 
We also note that points in the Argonne spectrum from about
E$_{\alpha}$=850 keV to 1050 keV are systematically higher
than the TRIUMF points. This, however, can be resolved by different
normalizations as the main peaks do not need to match perfectly.

\begin{figure}
\center
\hspace*{-0.5cm}
\includegraphics[width=6.5cm,angle=-90]{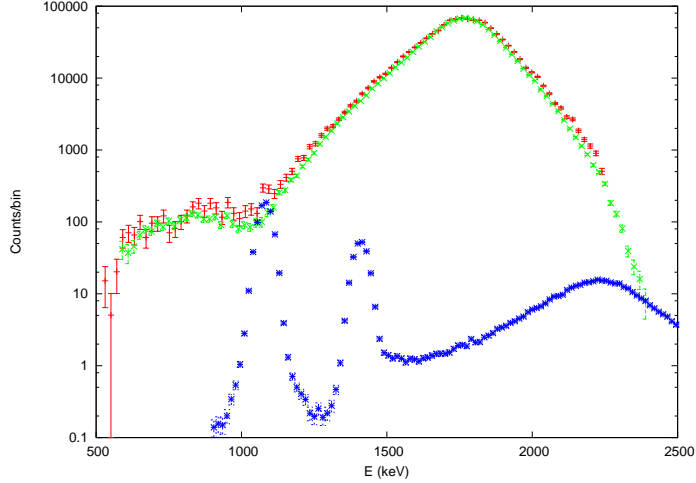}
\caption{(color online) Comparison of the $\beta$-delayed $\alpha$-spectra of $^{16}$N
Ref. \cite {Azu94} (x), Ref. \cite {Tan07} (bar) with the scaled $^{18}$N
spectrum of Ref. \cite {Buc07} (asterisks).}
\label {fig:argn18}
\end {figure}

Most interesting are the two points at about E$_{\alpha}$=1074 and 1094 keV.
In Fig. \ref {fig:argn18} we show the previous Figure \ref {fig:argcomp}
with the scaled $\beta$-delayed  $^{18}$N spectrum of Ref. \cite {Buc07}. 
No particular adjustment in the energy scale of either spectrum has been done.
The agreement with the main peak of the $^{18}$N spectrum for these
two points is remarkable. The lower energy point
is about 3.6$\sigma$ above its expected value and the higher energy
point 2.6$\sigma$. The statistical deviation has been determined
from R-matrix fits excluding this region, see Sec.
\ref {sec:rmatar}. The random chance for such a peak at the 
energy of the $^{18}$N main peak
is about 3$\times$10$^{-6}$. This strongly indicates a background of $^{18}$N
in the beam. While it is argued in Ref. \cite {Tan07} that the combination of
$^{15}$N+d cannot produce any $^{17,18}$N, fusion evaporation
reactions of the $^{15}$N beam with carbon or other light elements at windows or
collimators can produce both isotopes. 

\begin{figure}
\center
\hspace*{-0.5cm}
\includegraphics[width=6.5cm,angle=-90]{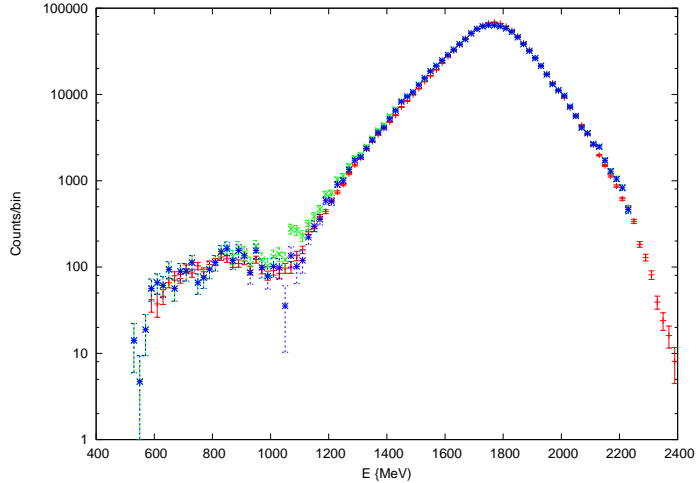}
\caption{(color online) Comparison of the $\beta$-delayed $\alpha$-spectra of $^{16}$N
of Ref. \cite {Tan07} (asterisk) with the one with $^{17,18}$N background removed
(cross), as described in the text. In addition, the TRIUMF spectrum is shown.}
\label {fig:argsub}
\end {figure}

While $^{18}$N is a relatively trivial contamination 
(see Sec. \ref {sec:subtr18N}) showing mainly up as one narrow peak and
being removable by decay time cuts, $^{17}$N is not. The $\beta$-delayed
$\alpha$ spectrum of $^{17}$N \cite {Dom94} has no narrow peak, and the
half life (4.2 s) is near that of $^{16}$N (7.1 s). Removing
both the likely $^{18}$N background and an $^{17}$N background of
roughly the same intensity leads to the spectrum shown in Fig. 
\ref {fig:argsub}. 
The spectrum derived by such a background subtraction agrees well
both on the plateau and
in the region above up to E$_{\alpha}$=1300 keV with the TRIUMF
spectrum. The low yield point at about E$_{\alpha}$=1054 keV is 
statistically consistent with the TRIUMF spectrum.
In its calculation, the method of the subtraction, 
and details like matching energy calibrations
and different detector resolutions that are involved pose 
rather subtle problems that are not worth improving at this stage.

\subsection {$R$-matrix fits to the Argonne spectrum}
\label {sec:rmatar}

\begin{figure}
\center
\hspace*{-0.5cm}
\includegraphics[width=6.5cm,angle=-90]{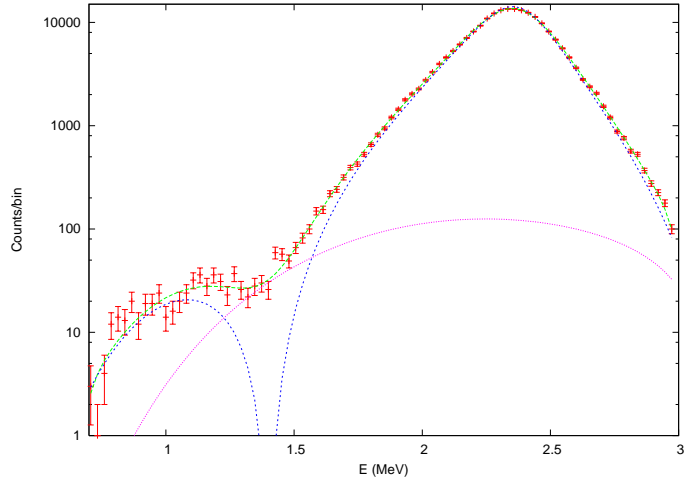}
\caption{(color online) $R$-matrix fits to the Argonne spectrum with $a$=5.5 fm. 
Shown are the spectrum (bar),
the resolution convoluted fit (long dash), the unconvoluted total fit 
(short dash), the p-wave component (dotted) and the f-wave component (dash-dotted).
The energy is in the center of mass system.}
\label {fig:arfit55}
\end {figure}

The Argonne spectrum has been fitted at different interaction
radii $a$. In general, we find the S-factors S$_{E1}$(300) to be
smaller than given in Ref. \cite {Tan07}. However, as our spectrum
is read out from a paper, we encounter large deviations from the
fit at the peak of the $^{16}$N distribution due to the difficulties
of reading out the spectrum there. While it is possible to adjust the points
to lower the least squares sum considerably, this may lead to biased
fits.  In Fig. \ref {fig:arfit55} we show a fit done with $a$=5.5 fm.
For the resolution, we chose 57 keV in the center-of-mass, very
close to the optimum (minimum $\chi^2$) setting. For this particular fit
S$_{E1}$(300) is found to be 68 keV b. As it is not described in 
Ref. \cite {Tan07} what constitutes a statistical error, we cannot derive those.
In addition, it is not mentioned what fit parameters have been used,
and what degrees of freedom are included in the partial least
squares parameters.

\begin{figure}
\center
\hspace*{-0.5cm}
\includegraphics[width=6.5cm,angle=-90]{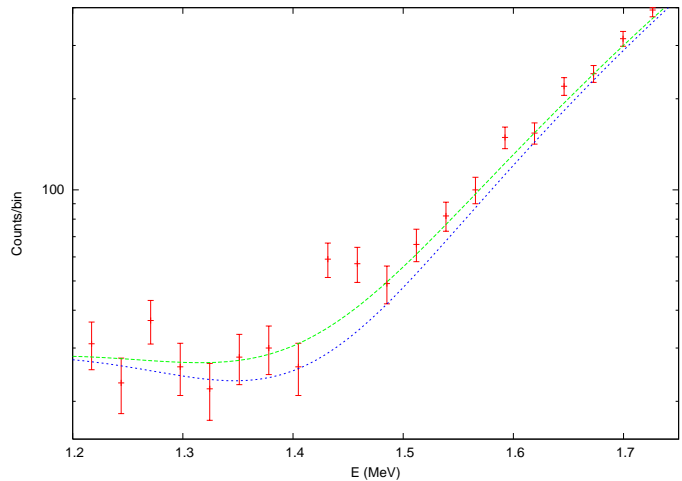}
\caption{(color online) $R$-matrix fits to the Argonne spectrum with $a$=5.5 fm
for the energy region displayed.
Shown are the spectrum (bar), the resolution convoluted fit 
(long dash), and the fit, with those data excluded (short dash).
The energy is the center of mass energy.}
\label {fig:arfit55noreg}
\end {figure}

The Argonne group has put great emphasis on the region between
E$_{\alpha}$=1.0 to 1.3 MeV showing a deviation from the TRIUMF data.
To see the relevance of this region, the data points there have been
taken out of the fit. Fig. \ref {fig:arfit55noreg} shows the fit in the region
of interest with the data used and not used in the fit.
Obviously, when those data are excluded, the fit runs close to the 
TRIUMF data. For the same exercise with the TRIUMF data the fit
does not change significantly. 

The deviation of those points identified as likely $^{18}$N 
events was taken from the fit that had those events removed,
see Sec. \ref {sec:comptarg}.

\subsection {Background subtraction in the TRIUMF spectrum}
\label {sec:back}

It has been suggested \cite {Fra07,Ade94} \footnote {Quoting from Ref. \cite
{Fra07}:`The disagreement could quite possibly be due to over subtraction
of $^{18}$N contamination in the TRIUMF data. [...] The disagreement in the
region of the interference minimum around 1.4 MeV is sufficient to change the 
f-wave component and leads to imprecise determination of the p-wave 
contribution to the $^{16}$N spectrum.' Obviously an imprecision beyond the
errors presented by the TRIUMF experimental group is meant.}
that the final TRIUMF result is questionable beyond the errors quoted, 
because a background of events from the 
$\beta$-delayed $\alpha$-decay of 
both $^{17}$N \cite {Dom94} and $^{18}$N \cite {Buc07} has
been subtracted to obtain the final spectrum. While the total number of these
subtracted events is only 2130 compared to 1,026,500 total events after subtraction, 
it has been implied that those
subtractions, if reversed, would (i) reconcile the differences between different 
measurements, and (ii) cause changes larger than the assigned errors given 
to the final S-factor S$_{E1}$(300) quoted in Ref. \cite {Azu94}.
We show below that these arguments are without merit, and could have been checked by the
authors of Refs. \cite {Fra07,Ade94} themselves, as both the total number
of events subtracted 
as well as the shape of these backgrounds are given in Ref. \cite {Azu94}, 
particularly in Fig. 10.

\subsubsection {The subtraction of the $\beta$-delayed $\alpha$-spectrum of $^{17}$N}

\begin{figure}
\center
\hspace*{-0.5cm}
\includegraphics[width=6.5cm,angle=-90]{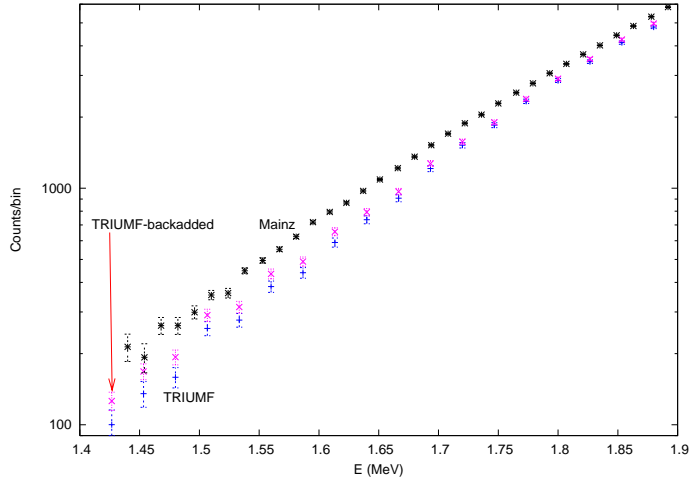}
\caption{(color online) Comparison of the $\beta$-delayed $\alpha$-spectra of $^{16}$N
Ref. \cite {Azu94} (bar) and Ref. \cite {Waf71}, (as energy-calibrated
in Ref. \cite {Fra07}) (asterisks) with the spectrum where previously subtracted
events from the  $\beta$-delayed $\alpha$-decay of 
$^{17}$N have been added back (cross). [Only the relevant low energy region is shown.]
The energy is the center-of-mass energy.}
\label {fig:n17comp}
\end {figure}

In Fig. \ref {fig:n17comp} we compare the final TRIUMF $^{16}$N spectrum, the 
TRIUMF $^{16}$N spectrum with all $^{17}$N events added back, 
and the Mainz spectrum (as energy calibrated by the Yale group \cite {Fra07}).
Clearly, the effect of the subtraction of the $^{17}$N spectrum from the
TRIUMF spectrum is minor, and does not produce any reasonable identity 
with the Mainz single event spectrum.
The S-factor at 300 keV is found to be S$_{E1}$(300)=82.2 keV b with
the $^{17}$N events added back to be compared with S$_{E1}$(300)=79$\pm$21 keV b
with the $^{17}$N events subtracted \cite {Azu94}. Note that in Ref. 
\cite {Azu94} an error of $\pm$5 keV b is given for the systematic uncertainty
in the R-matrix fits resulting from the $^{17}$N event subtraction.
It is therefore concluded that a claim of significant uncertainties resulting
from the subtraction of $^{17}$N events is irrelevant.

\subsubsection {The subtraction of the $\beta$-delayed $\alpha$-spectrum of $^{18}$N}

\label {sec:subtr18N}

The $\beta$-delayed $\alpha$ emitter $^{18}$N has a half life of 0.63 s. As reported
in Ref. \cite {Azu94}, the implantation time for the radioactive $^{16}$N beam was 3 s with
0.25 s moving time between three subsequent decay stations. While there is 
small but clear
evidence of the low energy 1.081 MeV peak from $^{18}$N in the spectra from the
first detector station, this peak is no longer visible in the subsequent two stations as
expected from the short $^{18}$N half life. The four later spectra were then used
to remove the $^{18}$N contamination from the $^{16}$N spectra. This information
has been given previously in Ref. \cite {Azu94}.

\begin{figure}
\center
\hspace*{-0.5cm}
\includegraphics[width=6.5cm,angle=-90]{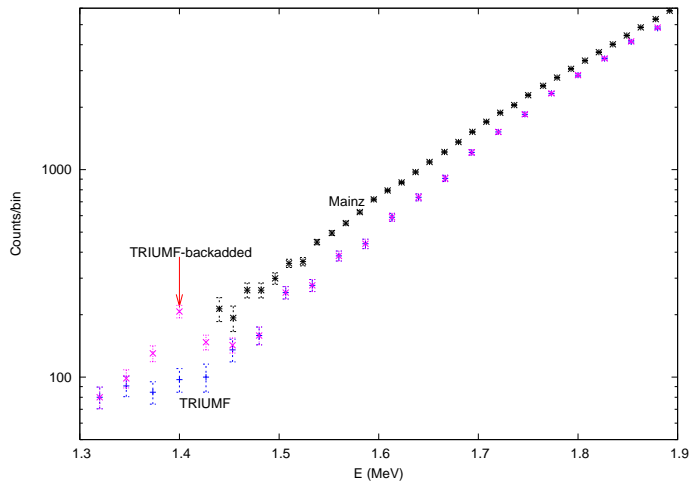}
\caption{(color online) Comparison of the $\beta$-delayed $\alpha$-spectra of $^{16}$N
Ref. \cite {Azu94} (bars) and Ref. \cite {Waf71} (asterisks), (energy calibrated
in Ref. \cite {Fra07}), with the 
TRIUMF spectrum to which all events from the  $\beta$-delayed $\alpha$-decay of 
$^{18}$N have been added back (cross). [Only the relevant low energy region is shown.]
The energy is in the center-of-mass system.}
\label {fig:n18comp}
\end {figure}

The low energy $^{18}$N peak
allows for an easy scaling of the $^{18}$N subtraction. 
In Fig. \ref {fig:n18comp} we compare the TRIUMF $^{16}$N spectrum, the 
TRIUMF $^{16}$N spectrum with all $^{18}$N events added back
and the Mainz spectrum, as calibrated by the Yale group \cite {Fra07}.
Again, it is clear that agreement with the Mainz spectrum is
not significantly improved by adding $^{18}$N events back. 
The fit to the added-back spectrum has
a relatively poor quality as it cannot accommodate the narrow $^{18}$N peak.
However, leaving the $^{18}$N events in the spectrum leads to an 
S$_{E1}$(300)=74.2 keV b, still well within the quoted error of 
Ref. \cite {Azu94}, i.e. 79$\pm$21 keV b. Note, however, that the $^{18}$N 
$\alpha$ peak in the $^{16}$N
spectrum provides an excellent intrinsic $\alpha$-particle energy calibration with a precision of
about 1 keV in this energy region.


\section {Conclusion}
\label {sec:conclusion}

In Table I we give an overview over the $^{16}$N decay
experiments discussed here.
\begin {table*}
\label {tab:over}
\caption {Overview of the $\beta$-delayed $\alpha$ decay experiments
          discussed in the article.}
\begin {tiny}          
\begin {tabular} {|ccccccc|} \hline
Experiment & Mainz \cite {Waf71} & TRIUMF \cite {Azu94} & Yale 1 \cite {Zha93} & 
                                            Seattle \cite {Ade94} & Yale 2 \cite {Fra07} & Argonne \cite {Tan07}\\ \hline
$^{16}$N production & $^{15}$N(d,p) & Isotope Separator & $^{15}$N(d,p) & $^{15}$N(d,p) & $^{15}$N(d,p) &  $^{15}$N(d,p)  \\
$^{16}$N implantation speed  &   low    &  low    &   high  &  high & high  & high     \\
mass/composition & 30 $\mu$g/cm$^2$C$_6$H$_7$N$_{2.5}$O$_{10}$ &   10 $\mu$g/cm$^2$ C&  180 $\mu$g/cm$^2$ Al & 10/20 $\mu$g/cm$^2$ C & 180 $\mu$g/cm$^2$ Al & 17 $\mu$g/cm$^2$ C  \\ 
of $^{16}$N catcher       &+0-1.5 cm 6-8 Torr $^{15}$N$_2$ gas &                     &                       &               &        & \\
                 & eqv. 0-17 $\mu$g/cm$^2$ N    &                     &                       &                  &     &  \\
   detectors     &  $<$35 $\mu$m Si  & 10.4-15.8 $\mu$m Si & 50 $\mu$m Si& 15-20 $\mu$m Si      &  50 $\mu$m Si & ion chamber\\       
   background (measured)   &$\beta$-tail &$^{18}$N, $^{17}$N &   none            &    unknown            &       none   & maybe \\
 degraded event suppression & none & $\alpha$-$^{12}$C &$\beta$-$\alpha$&  $\alpha$-$^{12}$C  &   $\beta$-$\alpha$ & $\alpha$-$^{12}$C \\
 efficiency corrections     & none  &   none          & $\beta$       &   unknown           &    $\beta$  & none     \\
 energy calibration & $^{10}$B(n,$\alpha$)& $^{18}$N, $^{20}$Na& $^{10}$B(n,$\alpha$)& unknown  & $^{10}$B(n,$\alpha$) &
  $^{10}$B(n,$\alpha$) \\
 deconvolution applied  &   none   &  none   & division (see text) & unknown &  division (see text) & none \\ \hline
\end {tabular}
\end {tiny}          
          
\end {table*}
Discussing the rows of Tab. I: Aside from the TRIUMF experiment, all others
 use the  $^{15}$N(d,p)$^{16}$N reaction for $^{16}$N production. However, only the
Mainz experiment \cite {Waf71}, the TRIUMF and the Argonne one 
thoroughly separated the region of $^{16}$N production from
the detection region. While care may have been taken to avoid $^{16}$N hitting e.g. 
the  catcher foil frames
in the Yale, Seattle and Argonne experiments,
a deep implantation of $^{16}$N ions has not been excluded at the level of 1/10,000 
leading to potentially degraded $\alpha$ events.

While the Mainz spectrum has been dubbed a `zero thickness spectrum' in Ref. \cite {Zha93}, it
is certainly not the one that uses the least amount of material to hold the $^{16}$N. As
the GEANT4 calculations show, the combination of an extended volume source
and the foils in front of the detectors leads to some broadening. In contrast,
the 10 $\mu$g/cm$^2$ C foils in the TRIUMF experiment 
slowly decreased in thickness in the course of the
implantations, until a hole was produced by the stable beam at mass A=30,
and the four active $^{16}$N collector foils had to be 
replaced. Spectra collected at different degrees of foil degradation
showed no difference. From Ref. \cite {Neu74} it is indeed known that for the Mainz spectrum 
 the accumulated mass traversed by
the $\alpha$-particles leads to an increase from about 15 keV detector resolution to
40 keV total resolution. The Mainz spectrum also allowed a relatively large
angular range of collection, leading to a slightly increased range in matter.
As far as the Seattle spectrum is concerned, it is not clear
whether the spectrum that has been distributed is from a 10 or 20 $\mu$g/cm$^2$ foil 
It also can not be excluded that additional carbon build up was involved. 
Of course, the catcher foils used by the Yale
group are much the thickest and, as shown above, this leads to heavily distorted spectra
dominated for low energies by scattering.

As far as detectors are concerned, all groups but one use silicon based detectors, with the
thinnest detectors, i.e. those with the least $\beta$ energy losses and thus $\beta$ 
response used by the TRIUMF group.
The thickness of the gas layer used to stop the $\alpha$-particles in the Argonne experiment
is comparable with the TRIUMF detectors.

Both, in the Mainz and the TRIUMF spectra, backgrounds have been subtracted. 
This issue is
extensively discussed above for the TRIUMF spectrum (Sec. \ref {sec:back}). 
In fact, for the Mainz spectrum, the 
$\beta$ induced background from the $^{16}$N decay is quite significant at low energies. 

As the primary goal of the Mainz experiment was not to determine the shape
of the $^{16}$N spectrum exactly, but to search for the narrow line of the parity-forbidden 
$\alpha$-decay of $^{16}$O at 1.283 MeV
of $\alpha$ energy, no measures were taken to check for degraded events. 
However, in all other experiments, some measures for increased discrimination 
against degraded or otherwise unwanted events were taken. In the case of the Yale
experiments $\beta$-$\alpha$ coincidences were employed. Those, in principle,
eliminate events from the $\beta$ tail of the $\alpha$-detector and from events, where 
$\alpha$ particles hitting the detector 
do not deposit the full charge. However, no means exists
in this method to eliminate events that suffer from some degradation of 
$\alpha$-particle energies in the thick aluminum foil. This is consistent with the 
observations and simulations about these measurements discussed above. 
As the detection of very low energy $\beta$-particles is required for 
high energy $\alpha$-particles, extensive $\beta$ efficiency corrections
were necessary, in addition, to derive the two Yale spectra.

In the Argonne, Seattle and TRIUMF experiments $\alpha$-$^{12}$C coincidences 
with subsequent pulse height ratio cuts were used.
That, in principle, removes $\beta$-tail events, a good fraction of 
degraded detector events and events scattered in the collection foil.
As our simulations show, a low level of degraded response events will remain
depending on catcher foil thickness.

While we do not know the method of energy calibration
for the Seattle experiment, the Argonne, Yale and Mainz 
experiments employed the $^{10}$B(n,$\alpha$)$^7$Li reaction
for their energy calibration. The reaction results in
two $\alpha$-particle groups, at E$_{\alpha}$=1472 and 1777 keV,
i.e. close together and close to the $^{16}$N main peak. One also does obtain
two lower energy $^7$Li groups. However, those are less suitable for calibration,
as different pulse height defects or other differences
for the differences in the detection of the $^7$Li ions are difficult
to take into account. Therefore, no low energy
calibration is available in these experiments. It should be noted that the method
of energy calibration is rather uncritical for the Yale experiments
as they are shifted upward by about 200 keV and use an invalid deconvolution method. 
In addition, Argonne used the $^6$Li($\alpha$,n)t reaction
for a high-energy calibration point.
The Argonne measurement
claims a smaller systematic error from their energy calibration (5\%) than
the TRIUMF one (10\%) \cite {Tan07b}. 

While not having employed it as a calibration in the TRIUMF experiment, we have observed
that the actual position of the $\alpha$ peaks from $^{10}$B(n,$\alpha$)$^7$Li
was not consistent with the calibration obtained
from $^{18}$N and $^{20}$Na. Indeed the calibration with $^{10}$B(n,$\alpha$)$^7$Li
 was found to be
dependent on the direction of the incident neutron flux, suggesting that only incomplete
neutron thermalization was achieved. Complete thermalization may be hard to prove
without an accurate measurement of the neutron spectrum at the position of the 
$^{10}$B target.

It is our opinion that the best energy calibration comes 
from the narrow decay lines of
well known $\beta$-delayed $\alpha$-emitters that can be used without modifying the 
experimental set-up.

Besides the Yale measurements, no other measurement relies on deconvolutions.
It has been demonstrated above that both methods applied are incorrect. It also has been
shown that both original measurements are consistent with the TRIUMF data 
when compared with simulations and, for the
case of France III \cite {Fra07}, inconsistent with their own `deconvoluted'
spectrum.

The TRIUMF spectrum, among all measurements, shows the 
narrowest main peak. It has been shown above that the energy calibration
used for the derivation of this spectrum is the best one available, and that the
shape of the TRIUMF spectrum is little influenced by background subtractions. 
In principle, besides resolution broadening, also an energy dependent efficiency
can lead to a broadening, and more likely narrowing (efficiency dropping off at low
energies) of the spectrum.
However, it is shown experimentally in Ref. \cite {Azu94} that the coincidence detection
efficiency stays constant over the energy range covered in the measurement. 
None of the other measurements considers this point. As far as broadening by target and
detector effects is concerned
to the authors of this article there is no reasonable physical response 
function known that would
transform the wider spectra of the other measurements into a narrower one, as
measured at TRIUMF. But there are many well known transformations that can produce
a wider spectrum from a narrow one.


\section {Outlook}
\label {sec:outlook}

We have presented fits to several spectra in this article. However, either the derivation of 
those spectra is in doubt, even if one does not
take systematic errors into account, as for the Yale spectra, or the spectra are not
completely published, as for the Argonne spectrum. Therefore any derivation of a 
common S-factor S$_{E1}$(300) would lead to questionable results. 
We therefore reiterate the 
value of S$_{E1}$(300) derived in Ref. \cite {Azu94}, i.e. S$_{E1}$(300)=79$\pm$21 keV b
including systematic errors.
A new evaluation of the entire reaction rate of 
$^{12}$C($\alpha$,$\gamma$)$^{16}$O is in work \cite {Buc08}
taking new information from other measurements into account, particularly phaseshift and
radiative capture data.

While other partial cross sections of $^{12}$C($\alpha$,$\gamma$)$^{16}$O have relative
uncertainties exceeding those of the $E1$ ground state transition, in the long run, it
certainly would be desirable to improve the value derived in Ref. \cite {Azu94} to
a relative error of less than 10\%. While statistics plays some role, experimental 
limitations and systematic errors are the real limits.

Some numbers from other measurements are part of the fit to the $\beta$-delayed
$\alpha$-spectrum of $^{16}$N. These are the energy of the subthreshold 1$^-$ state,
its radiative width, and the $\beta$-decay branching ratio into the 
E$_x$=9.6 MeV 1$^-$ state in $^{16}$O. While the former numbers appear reasonably well
known, there is some uncertainty about the latter branching ratio. The compilation
of Ref. \cite {Tun93} quotes 1.20$\times$10$^{-5}$$\pm$0.05 for 
this branching ratio citing Ref. \cite {Cho93}
as the origin of this number. However, Ref. \cite {Cho93} is a theoretical work quoting
Ref. \cite {Bar71}, another theoretical work. Ref. \cite {Bar71} quotes Ref. 
\cite {Hat70} (an early publication of the Mainz group) 
which gives a branching ratio of 1.19$\times$10$^{-5}$$\pm$0.10. 
Both theoretical works do not quote an experimental error, so the origin of the
error in Ref. \cite {Tun93}is unknown to us. However, Ref. \cite {Neu74}, p. 327,
gives a revised value of 1.13$\times$10$^{-5}$$\pm$0.08 for the branching
ratio into the E$_x$=9.6 MeV state of $^{16}$O, a not insignificant 
deviation. Therefore a revised measurement with improved errors
of this branching ratio is desirable.

Among the errors inherent to the measurement is the energy calibration of the 
spectrum. Ref. \cite {Azu94} quotes a systematic error of 10 keV b for S$_{E1}$(300),
while Ref. \cite {Tan07b} quotes only 5 keV b with a more indirect measurement. 
This error can certainly be improved.

Part of the uncertainties in the fits originates from the presence of both a $p$- as well
as an $f$-wave in the $\beta$-delayed $\alpha$-spectrum of $^{16}$N. Higher statistics
data at very low energies (E$_{\alpha}<$ 600 keV) would be very desirable in this context
as they originate only from $p$-wave decay, while due to penetrability reasons the
$f$ wave drops off faster and at higher energies. However, this $\alpha$-energy 
region is difficult to access, first
because of possible $\beta^-$ back ground in the detectors that eventually 
may be removed, but more important
second because of a possible response of high energy $\alpha$ events into this region.
Our simulations have shown (Secs. \ref {sec:seattlesim}, \ref {sec:argonnesim})
that indeed the most advanced experiments using carbon foils
of about 10 $\mu$g/cm$^2$ and coincidence techniques 
are close to this response limit. Foils with much less 
mass seem to be impractical. As an alternative $^{16}$N 
could be trapped in electromagnetic or
opto-magnetic traps with no substrates for the $^{16}$N involved. Because of 
trapping efficiencies and likely difficult geometries such an experiment will require
high yields of $^{16}$N. In this case careful simulations of the next order of response
like those originating from the detectors are necessary. Another possibility, in principle,
is to measure the $\beta$-$\alpha$ correlations in the decay of polarized $^{16}$N that
would differentiate the data between the $p$- and the $f$-wave in the  region
of about and above E$_{\alpha}$=1 MeV, see Ref. \cite {Ji90}.

We wish to thank Prof. C.A. Barnes of the Kellogg Radiation Laboratory, California
Institute of Technology, for fruitful discussions. We also received comments about 
the present manuscript from
Prof. J. D'Auria, Simon Fraser Unversity, Prof R.E. Azuma (University
of Toronto), Prof. J. King (U. of Toronto), Dr. P. Mc Neely (MPI Plasmaphysik),
Dr. M. Dombsky (TRIUMF), Dr. J. Powell (Lawrence Berkeley Laboratory),
and Dr. P. Wrean (Camosum College). 


\end {document}